\documentclass[prd,twocolumn,showpacs,floatfix,amsmath,nofootinbib,amssymb,floatfix]{revtex4}
\usepackage{graphicx,color,dcolumn,booktabs,bm}
\usepackage{longtable,lscape}
\usepackage{txfonts}
\usepackage{overpic}
\usepackage{amssymb}
\usepackage{indentfirst}
\usepackage{feynmf}   
\usepackage{slashed}  
\usepackage{cases}
\usepackage{color}
\usepackage{multirow}
\usepackage{epstopdf}
\usepackage{longtable}
\usepackage{graphicx,color,dcolumn,booktabs,bm}
\usepackage[colorlinks,
            citecolor=blue,
            anchorcolor=red,
            menucolor=red,
            linkcolor=red,
            filecolor=red,
            runcolor=red,
            urlcolor=blue,
            frenchlinks=red]{hyperref}

\graphicspath{{Figures/}} %

\allowdisplaybreaks

\begin{document}

\title{Regge Behaviors in Orbitally-excited Spectroscopy of Charmed and
Bottom Baryons}

\author{Duojie Jia$^{1,2}$}
\email{jiadj@nwnu.edu.cn}
\author{Wen-Nian Liu$^{1}$}

\author{Atsushi Hosaka$^{2}$}

\affiliation{ $^1$Institute of Theoretical Physics, College of Physics and
Electronic Engineering, Northwest Normal University, Lanzhou 730070, China
\\
$^2$Research Center for Nuclear Physics, Osaka University,
10-1 Mihogaoka, Ibaraki, Osaka 567-0047, Japan}

\begin{abstract}
Stimulated by recent progress made by the LHCb Collaboration in discoveries
of new bottom baryons, e.g., the $\Xi _{b}(6227)^{-}$ and the $\Sigma
_{b}(6097)^{\pm}$, we re-examine the orbitally excited spectrum of the charmed
and bottom baryons using Regge approach in the heavy quark-diquark picture.
The results indicate that the spin-averaged mass spectrum of the
orbitally-excited charmed and bottom baryons can be described by a linear
Regge relation, which is derived from the rotating QCD string model. By giving further mass-splitting analysis of spin-dependent interactions, we
explain the baryons $\Xi_{b}(6227)^{-}$ and the $\Sigma_{b}(6097)^{\pm}$,
and the $\Sigma _{c}(2800)$ and $\Xi _{c}^{\prime }(2930)$ to be the $1P$%
-wave baryons, all with the spin-parity $J^{P}=3/2^{-}$ preferably. Mass prediction of the bottom baryon $\Xi_{b}$ in its P- and D-waves are presented, providing clues for the coming experiments like the LHCb to find them.
\end{abstract}


\maketitle

\section{Introduction}\label{sec1}

The information from singly heavy (SH) baryon spectrum is quite helpful
to deepen the understanding of non-perturbative Quantum Chromodynamics (QCD)
as the heavy quark inside a baryon endows us with a `flavor tag' to look
further into the nature of the interquark forces \cite{IsgurWise:L91}. This forms an analogy with the hydrogen-like atoms, in which the quantum states of the binding systems are labeled well by that of light degrees of freedom, and provides us with a good chance to further construct excited states of the
bottom baryon family. In the past two decades, more than twenty charmed baryon candidates have been announced by the different experimental collaborations
\cite{Tanabashi:D18}. Differing from the charmed baryons, searching for the bottom baryon states is far from easy for experiment since the bottom baryons
seems to be more difficult to produce than the charmed baryons. Only several bottom baryon candidates were reported before 2017.

Recently, the LHCb Collaboration discovered, one after another, two new
bottom baryon states, the $\Xi _{b}(6227)^{-}$ by studying its final states
of the $\Lambda _{b}^{0}K^{-}$ and $\Xi _{b}^{0}\pi ^{-}$ \cite{LHCb:Xib2018}%
, and the $\Sigma _{b}(6097)^{\pm }$ in the final states of $\Lambda
_{b}^{0}\pi ^{\pm }$ \cite{LHCb:Xib2019}. The two baryons differ
about $130$MeV in mass, close to the mass difference of $u/d$ and $s$
constituent quarks, and exhibit similar decay behaviors in width and
final states. After their observations, the two baryons become the key states for establishing the whole family of the excited bottom baryons, and their properties and inner structures become hot topics for the heavy baryon study\cite{AASS:D2018,CL:D2018,CWLZ:D2018,PYang:D2019,WangLZ:D19,Cui:2019dzj,Aliev:2018vye}. Most of these studies prefer to suggest that the $\Xi_{b}(6227)^{-}$ is the $P$-wave strange partner of the $\Sigma _{b}(6097)^{\pm}$, with spin-parity $J^{P}=3/2^{-}$ or $5/2^{-}$ while there are also the molecular accommodation\cite{HuangXG:D19}. For more discussions on the excited heavy baryons, see Refs.\cite{KornerKP:P94,CredeR:D13,HYCheng:F15,QMao,EFG:D11,EFG:B08,KlemptR:RM10,RobertsP:A08}.

To further decode inner structure of the $\Xi _{b}(6227)^{-}$ and $\Sigma
_{b}(6097)^{\pm }$, it becomes important to explore the charmed and bottom
baryons in a coherent way so that these newly discovered baryons can be
analysed in connection with the existing singly heavy(SH) baryons in the
charm sector, e.g., the $\Sigma _{c}(2800)$ and $\Xi _{c}^{\prime }(2930)$,
whose quantum numbers are not established experimentally yet
\cite{Tanabashi:D18,Aubertet:D08,Li:C18}. Obviously, it will be helpful to deepen
our understanding of non-perturbative QCD if some analogies were found among these SH hadron systems. For the two charmed baryons mentioned above, quark models
predict their quantum number to be \cite{EFG:D11,EFG:B08,KlemptR:RM10,RobertsP:A08},
\begin{quote}
\begin{equation}
J^{P}(\Sigma _{c}(2800))=\frac{1}{2}^{-}\text{ or }\frac{3}{2}^{-},J^{P}(\Xi
_{c}^{\prime }(2930))=\frac{3}{2}^{-}\text{or }\frac{5}{2}^{-}.  \label{JP}
\end{equation}
\end{quote}
Based on the Regge trajectories, authors of Ref.\cite{ChengC:D17} assign the $\Sigma _{c}(2800)$ and $\Xi_{c}^{\prime }(2930)$ to be the $P$-wave states with $J^{P}=3/2^{-}$.

The purpose of this work is to explore the orbitally-excited mass spectrum of the
SH baryons $qqQ$ with the help of Regge trajectory and accommodate
the newly discovered bottom baryons $\Xi _{b}(6227)$, the $%
\Sigma_{b}(6097)^{\pm}$, as well as the charmed baryons $\Sigma_{c}(2800)$
and $\Xi_{c}^{\prime}(2930)$. Our mass analysis indicates that a linear Regge
trajectory is able to describe the spin-averaged mass spectrum of the
existing singly heavy baryons in their orbitally excited states. Further mass-splitting analysis of the orbitally excited SH baryons suggests that the
quantum numbers of the four SH baryons considered here are $J^{P}=3/2^{-}$ preferably. We also make the mass prediction of the bottom baryon $\Xi_{b}$ in its P- and D-waves, which should be tested by future experiments, especially the LHCb.

The paper is organized as follows. After the introduction, we show how the
linear Regge relation can be employed to demonstrate Regge behaviors in
the spin-averaged masses of orbitally excited SH baryons in Sec. II, where the trajectory parameters are discussed. In sec. III, we perform mass calculations for the orbital excitations of the SH baryons with $L=0,1$ and $2$($\Lambda _{c}$,$\Xi _{c}$ cases) by including the spin-dependent interactions, and assign the $J^{P}$ quantum number for the $%
\Sigma _{c}(2800)$ and the $\Xi _{c}^{\prime }(2930)$. In Sec. IV the similar
analysis is applied to the singly bottom baryons and to assign the quantum number
for the newly reported baryons $\Xi _{b}(6227)$ and the $%
\Sigma_{b}(6097)^{\pm}$. The masses of the bottom baryon states $\Xi _{b}$ in the P- and D-waves are also predicted. Finally, the paper ends
with the discussions and summary in Sec. V.

\section{Spin-independent masses for singly heavy baryons}\label{sec2}

The notion of Regge trajectory or Regge mass relation $L=\alpha^{\prime }M^{2}+\alpha_{0}$ for hadrons, which connects high energy scattering and the spectrum of hadrons in general theory \cite{Regge:59}, provides us a powerful way to organize the whole spectrum of a given family of the hadrons. In the case of light hadrons, the Regge trajectories are commonly believed to be linear and parallel, and these features are also tested to be true roughly for the SH hadrons in Ref. \cite{KChenL:C18} provided that the hadron mass undergoes a shift $M$ $\rightarrow$ $M-M_{Q}$, with $M_{Q}$ the heavy quark mass. As such, one can estimate the spin-averaged masses of $\Sigma _{Q}$ and $\Xi _{Q}^{\prime }$ in their orbitally-excited states with the help of the trajectory information of the partner SH baryons or mesons.

We begin with a linear Regge relation \cite{BChen:A15,ChenCL:Rpp17} , which is derived from the rotating string model (see Appendix A),
\begin{equation}
\left( M-M_{Q}\right) ^{2}=\pi aL+\left( m_{d}+\frac{P_{Q}^{2}}{M_{Q}}%
\right) ^{2},  \label{GRegg}
\end{equation}%
This relation connects the shifted mass of a SH baryon squared, $\left(M-M_{Q}\right) ^{2}$, with its quantum number of orbital angular momentum $L$. Here, $a$ stands for the tension of the string connected to the heavy quark with mass $M_{Q}$ at one end and the light diquark with mass $m_{d}$ at the other, and $M$ is the mass of the heavy baryon. In Eq.(\ref{GRegg}), the intercept depends on the diquark mass $m_{d}$ and the non-relativistic kinematic energy
$P_{Q}^{2}/M_{Q}$ of the heavy quark through $a_{0}=\left(m_{d}+P_{Q}^{2}/M_{Q}\right) ^{2}$, with
\begin{equation}
P_{Q}\equiv M_{Q}v_{Q}\simeq M_{Q}\left( 1-\frac{m_{bareQ}^{2}}{M_{Q}^{2}}%
\right) ^{1/2},  \label{PQ}
\end{equation}%
where $m_{bareQ}=1.275$\ \text{GeV} and $4.18$\ \text{GeV} are the bare masses of the heavy charmed quark $Q=c$ and bottom quark $Q=b$, respectively. Here, $v_{Q}$=$\left( 1-\frac{m_{bareQ}^{2}}{M_{Q}^{2}}\right) ^{1/2}$ stands for the velocity of the heavy quark, which is conserved in the heavy quark limit.

The simplest way to understand Eq. (\ref{GRegg}) is to view it as an
extension of the Selem-Wilczek relation \cite{SelemW06} for the SH hadrons,
\begin{equation}
M-M_{Q}=\sqrt{\frac{\alpha }{2}L}+2^{1/4}\kappa \frac{\mu ^{3/2}}{L^{1/4}},
\label{SW}
\end{equation}%
which scales as $(M-M_{Q})^{2}\propto (\alpha /2)L$ when $L\rightarrow
\infty $ but divergent at $L=0$. As $L\gg 1$, Eq.(\ref{GRegg}%
) reduces to Eq.(\ref{SW}), $M-M_{Q}=\sqrt{\pi aL}+a_{0}/(2\sqrt{\pi aL%
})$, while it is finite at $L=0$ and becomes
\begin{equation}
M(1S)=M_{Q}+m_{d}+\frac{P_{Q}^{2}}{M_{Q}}\text{. }  \label{MS}
\end{equation}%
This is consistent with the prediction by the heavy quark symmetry for the ground state ($L=0$) \cite{Manohar:D07} as $M_{Q}$ is very close to the bare mass of the heavy quark $Q$ when $Q$ is heavy, as indicated by Eq.                                                                             (A3) in Appendix A. The heavy quark limit defined by $m_{bareQ}\rightarrow\infty$ implies $M_{Q}\simeq m_{bareQ}\rightarrow\infty$ and the 3-velocity of heavy quark $v_{Q}\rightarrow 0$(see Appendix A). Here, the $m_{d}$ plays the role of the energy of the light brown muck in a baryon in the lowest states. We note that the orbital angular momentum $L$ here corresponds to the $\lambda$ mode in the quark model \cite{Copley:1979wj,Nagahiro:2016nsx}.

In the light of the rotating string model \cite{SelemW06}, one can view the heavy baryon mass $M$ as the sum of the mass $M_{Q}$ of heavy quark at one end of the string and the string energy $\sqrt{\pi aL+a_{0}}$, with the light diquark tied to the other. We note that the full quantum treatment \cite{Rebbi:PR74,SW:jh18} of the QCD string provides merely an additional constant $\alpha_{0}$ added to $L$ as expected by the Regge trajectory of hadrons.

As will be seen in what follows, the mass of the diquark $(qq)_{d}$ in Eq. (\ref{GRegg}) varies when the strangeness and its spin involve. In a SH baryon with zero strangeness, the diquark can be either of scalar (spin$=0$) or of vector (spin=$1$),
\begin{equation*}
\text{ }(qq)_{d}=\left\{
\begin{array}{ll}
\lbrack ud], & I=0=\text{spin, in }\Lambda _{Q}, \\
\{uu,ud,dd\}, & I=1=\text{spin, in }\Sigma _{Q}.%
\end{array}%
\right.
\end{equation*}%
while in those with strangeness$=-1$, it can be either of scalar (spin%
$=0$) or axial vector (spin=$1$)

\begin{equation*}
\text{ }(qs)_{d}=\left\{
\begin{array}{ll}
\lbrack us,ds], & I=1/2,\text{spin}=0\text{, in }\Xi _{Q}, \\
\{us,ds\}, & I=1/2,\text{spin}=1\text{, in }\Xi _{Q}^{\prime }.%
\end{array}%
\right.
\end{equation*}%

To confront Eq.(\ref{GRegg}) with the baryon mass data, we list in Table I the observed masses of the heavy baryons $\Lambda _{Q}$ and the strange heavy baryons $\Xi _{Q}$ ($Q=c,b$), and in Table II that of the heavy baryons $\Sigma _{Q}$ and the strange heavy baryons $\Xi _{Q}^{\prime }$ ($Q=c,b$). In Tables IX, the observed spin-averaged masses for the family of orbitally excited SH baryons are collected with the weight $2J+1$(with $J$ the angular momentum of the baryon), and are fitted by the relation (\ref{GRegg}), in which the estimated mass can be rewritten as, upon using Eq. (\ref{PQ}),
\begin{equation}
M_{L}=M_{Q}+\sqrt{\pi aL+\left[m_{d}+M_{Q}-m_{bareQ}^{2}/M_{Q}\right] ^{2}}.
\label{MnL}
\end{equation}

\renewcommand\tabcolsep{0.53cm}
\renewcommand{\arraystretch}{1.5}
\begin{table*}[!htbp]
\caption{The observed quantum numbers and masses (in MeV) \cite{Tanabashi:D18} of the charmed and charm-strange baryons $\Lambda _{c}$
 and $\Xi _{c}^{0}$ that contain the scalar-diquark. The $J^{P}$ of some states indicated by the question marks are the quark-model predictions \cite%
{ChenCL:Rpp17}, and have not been established experimentally. The errors less
than $5$ MeV are not indicated.}\label{Table I}
\begin{tabular}{c|ccc|ccc}
\hline\hline
States, $J^{P}$ & Baryon & Mass & This work & Baryon & Mass & This work
\\ \hline
$1^{2}S_{1/2}, 1/2^{+}$ & $\Lambda _{c}^{+}$ & $2286.46(14)$ & $2286.0$ & $\Xi _{c}^{+}$ & $2467.87(30)$ & $2469.1$
\\
$1^{2}P_{1/2}, 1/2^{-}$ & $\Lambda _{c}(2595)^{+}$ & $2592.25(28)$ & $2588.7$ & $\Xi _{c}(2790)^{+}$ &
 $2792.0(5)$ & $  2778.6 $
\\
$1^{2}P_{3/2}, 3/2^{-}$ & $\Lambda _{c}(2625)^{+}$ & $2628.11(19)$ & $ 2628.9 $ & $\Xi _{c}(2815)^{+}$ &
$2816.67(31)$ & $ 2816.5 $
\\
$1^{2}D_{3/2}, 3/2^{+}$ & $\Lambda _{c}(2860)^{+}$ & $2856.1_{-6.0}^{+2.3}$ & $ 2857.3 $ & $\Xi_{c}(3055)^{+} $ & $ 3055.9(4)$ & $ 3058.7 $
\\
$1^{2}D_{5/2}, 5/2^{+}$ & $\Lambda _{c}(2880)^{+}$ & $2881.63(24)$ & $ 2880.2 $ &
$\Xi _{c}(3080)^{+}$ & $3077.2(4)$ & $ 3079.7 $
\\ \hline
$1^{2}S_{1/2}, 1/2^{+}$ & $\Lambda _{b}^{0}$ & $5619.60(17)$ & $5615.5$ & $\Xi _{b}$ & $5791.9(5)$ & $5792$
\\
$1^{2}P_{1/2}, 1/2^{-}$ & $\Lambda _{b}(5912)^{0}$ & $5912.20(21)$ & $ 5908.5 $ & $ \Xi _{b} $ &   & $6116.9$
\\
$1^{2}P_{3/2}, 3/2^{-}$ & $\Lambda _{b}(5920)^{0}$ & $5919.92(19)$ & $ 5921.4 $  & $ \Xi _{b} $ &  & $6129.1$
\\
$1^{2}D_{3/2}, 3/2^{+}$ & $\Lambda _{b}(6146)^{0}$ & $6146.17$ & $ 6144.8 $ & $ \Xi _{b} $ &  & $6376.9$
\\
$1^{2}D_{5/2}, 5/2^{+}$ & $\Lambda _{b}(6152)^{0}$ & $6152.51$ & $ 6152.2 $  & $ \Xi _{b} $ &  & $6383.6$
\\ \hline\hline
\end{tabular}
\end{table*}
\medskip

\renewcommand\tabcolsep{0.53cm}
\renewcommand{\arraystretch}{1.8}
\begin{table*}[!htbp]
\caption{The observed quantum numbers and masses (in MeV) \cite{Tanabashi:D18} charmed and charm-strange baryons that contain vector-diquark. The $J^{P}$ of some states indicated by the question marks are the
quark-model predictions.}\label{Table II}
\begin{tabular}{c|ccc|ccc}
\hline\hline
States, $J^{P}$ & Baryon & Mass & This work & Baryon & Mass & This work \\ \hline
$1^{2}S_{1/2}, 1/2^{+}$ & $\Sigma _{c}(2455)^{++}$ & $2453.97(14)$ & $ 2452.7 $& $\Xi _{c}^{\prime }$ & $2578.8(5)$ & $  2586.0 $
\\
$1^{4}S_{3/2}, 3/2^{+}$ & $\Sigma _{c}(2520)^{++}$ & $2518.41_{-0.19}^{+0.21}$ & $  2517.8 $ & $\Xi _{c}^{\prime 0}(2645)$ & $2646.32(31)$& $2641.6$
\\
$1^{2J+1}P_{J}, ?^{?}$ & $\Sigma _{c}(2800)^{++}?$ & $2801_{-6}^{+4}$ &$--$ & $\Xi _{c}^{\prime }(2930)?$ & $2931(6)$ & $--$
\\ \hline
$1^{2}S_{1/2}, 1/2^{+}$ & $\Sigma _{b}^{+}$ & $5811.3(1.9)$ & $ 5811.0 $& $\Xi _{b}^{\prime }(5935)^{-}$ & $5935.02(05)$ & $5937.1$
\\
$1^{4}S_{3/2}, 3/2^{+}$ & $\Sigma _{b}^{*+}$ & $5832.1(1.9)$ & $  5832.0 $& $\Xi _{b}^{\prime }(5955)$ & $5955.33(13)$& $5955.0$
\\
$1^{2J+1}P_{J}, ?^{?}$ & $\Sigma _{b}(6097)^{+}?$ & $6095.8(2)$ & $--$ & $\Xi _{b}^{\prime }(6227)?$ &
$6226.9(2.1)$ & $--$
\\ \hline\hline
\end{tabular}
\end{table*}
\medskip
\medskip

\renewcommand\tabcolsep{0.64cm}
\renewcommand{\arraystretch}{1.8}
\begin{table*}[!htbp]
\caption{The effective masses (GeV) of the charm quark and scalar-diquarks, and the tension (GeV$^{2}$) that match the measured spin-averaged masses of the ${\small \Lambda_{c}}$ and the ${\small\Xi_{c}}$ in Table I. Here, the RMS error $\chi_{RMS}=0.001$GeV and $n=u/d$. The comparison with that by quark model is given. }\label{Table III}
\begin{tabular}{c|ccccccccc}
\hline\hline
{\small Parameters} &{${\small M}_{c}$} & ${\small m}_{d}{\small ([nn])}$ & ${\small a(\Lambda }_{c}{\small )}$ & $%
{\small m}_{d}{\small ([ns])}$ & ${\small a(\Xi }_{c}{\small )}$ & ${\small \bar{M}(\Lambda _{c})}$ & ${\small \bar{M}(\Xi _{c})}
$ \\ \hline
{\small This work} & ${\small 1.44}$ & ${\small 0.535}$ & $%
{\small 0.212}$ & ${\small 0.718}$ & ${\small 0.255}$ & $%
{\small 2.618}$ & ${\small 2.804}$ \\
{\small EFG}\cite{EFG:D11} & {${\small 1.55}$} & ${\small 0.710}$ & ${\small 0.18}$ & ${\small 0.948}$ & ${\small %
0.18}$ & ${\small 2.617}$ & ${\small 2.808}$ \\ \hline\hline
\end{tabular}
\end{table*}
\medskip

\renewcommand\tabcolsep{0.59cm}
\renewcommand{\arraystretch}{1.8}
\begin{table*}[!htbp]
\caption{The effective masses (GeV) of the charm quark and vector-diquarks, and the tension (GeV$^{2}$) that match the measured spin-averaged masses of the $\Sigma_{c}$ and the $\Xi_{c}^{\prime}$ in Table II. Here, the RMS error $\chi_{RMS}=0.001$GeV and $n=u/d$. The comparison with that by quark model is given.}\label{Table IV}
\begin{tabular}{c|cccccccc}
\hline\hline
{\small Parameters} & {${\small M}_{c}$} & ${\small m}_{d}({\small \{nn\}})$ & ${\small a}({\small \Sigma }%
_{c})$ & ${\small m}_{d}{\small (\{ns\})}$ & ${\small a(\Xi }_{c}^{\prime }%
{\small )}$ & ${\small \bar{M}(\Sigma _{c})}$ & $%
{\small \bar{M}(\Xi _{c}^{\prime})}$ \\ \hline
{\small This work} & ${\small 1.44[\text{input}]}$ & ${\small 0.745}$ & ${\small 0.212}$ & $%
{\small 0.872}$ & ${\small 0.255}$ & ${\small 2.774}$ & ${\small 2.923}$
\\
{\small EFG}\cite{EFG:D11} & {${\small 1.55}$} & ${\small 0.99}$ & ${\small 0.18}$ & ${\small 1.069}$ & ${\small 0.18%
}$ &  ${\small 2.780}$ & ${\small 2.919}$ \\ \hline\hline
\end{tabular}
\end{table*}
\renewcommand\tabcolsep{0.64cm}
\renewcommand{\arraystretch}{1.8}
\begin{table*}[!htbp]
\caption{The masses (GeV) of the bottom quark and diquarks, and the tension (GeV$^{2}$) that match the measured spin-averaged masses of the ${\small \Lambda _{b}}$ and the ${\small \Xi_{b}}$ in Table I and the $ {\small\Sigma_{b}}$ and the ${\small \Xi_{b}^{\prime}}$ in Table II. Here, the RMS error $\chi_{RMS}=0.001$GeV. The comparison with that by quark model is given.}\label{Table V}
\begin{tabular}{c|cccccccc}
\hline\hline
{\small Parameters} & ${\small M}_{b}$ & ${\small m}_{d}{\small ([nn])}$ & $%
{\small a(\Lambda }_{b}{\small )}$ & ${\small m}_{d}{\small ([ns])}$ & $%
{\small a(\Xi }_{b}{\small )}$  & ${\small \bar{M}%
(\Lambda _{b})}$ & ${\small \bar{M}(\Xi _{b})}$ \\ \hline
{\small This work} & ${\small 4.48}$ & ${\small 0.534}$ & ${\small 0.246}$ &
${\small 0.718}$ & ${\small 0.307}$ & ${\small 5.917}$ & $%
{\small 6.125}$ \\
{\small EFG}\cite{EFG:D11} & ${\small 4.88}$ & ${\small 0.710}$ & ${\small %
0.18}$ & ${\small 0.948}$ & ${\small 0.18}$ & ${\small 5.938}$ & ${\small %
6.127}$ \\ \hline\hline
\end{tabular}
\end{table*}
\medskip
\renewcommand\tabcolsep{0.59cm}
\renewcommand{\arraystretch}{1.8}
\begin{table*}[!htbp]
\begin{tabular}{c|cccccccc}
\hline\hline
{\small Parameters} & ${\small M}_{b}$ & ${\small m}_{d}({\small \{nn\}})$ &
${\small a}({\small \Sigma }_{b})$ & ${\small m}_{d}{\small (\{ns\})}$ & $%
{\small a(\Xi }_{b}^{\prime }{\small )}$ & ${\small
\bar{M}(\Sigma _{b})}$ & ${\small \bar{M}(\Xi _{b}^{\prime })}$ \\ \hline
{\small This work} & ${\small 4.48[\text{input}]}$ & ${\small 0.745}$ & ${\small 0.246}$ & ${\small %
0.869}$ & ${\small 0.307}$ & ${\small 6.088}$ & ${\small 6.248}$ \\
{\small EFG}\cite{EFG:D11} & ${\small 4.88}$ & ${\small 0.909}$ & ${\small %
0.18}$ & ${\small 1.069}$ & ${\small 0.18}$ &  ${\small 6.090}$ & ${\small 6.228}$ \\
\hline\hline
\end{tabular}
\end{table*}
\medskip
\renewcommand\tabcolsep{0.5cm}
\renewcommand{\arraystretch}{1.5}
\begin{table*}[!htbp]
\caption{The observed masses (in MeV) of the charmed and charmed strange
mesons \cite{Tanabashi:D18}. The some of quantum numbers indicated by
question marks are quark model predictions, which has not been established
experimentally. The errors less than $5$ MeV are not indicated.}\label{dm}
\begin{tabular}{ccccccc}
\hline\hline
{\small State }$J^{P}$ & {\small Meson} & {\small Mass} &
{\small EFG \cite{EFG:C10}} & {\small Meson} & {\small Mass} &
{\small EFG \cite{EFG:C10}} \\
\hline
$%
\begin{array}{rr}
{\small 1}^{1}{\small S}_{0} & {\small 0}^{-} \\
{\small 1}^{3}{\small S}_{1} & {\small 1}^{-}%
\end{array}%
$ & $%
\begin{array}{r}
{\small D}^{\pm } \\
{\small D}^{\ast }{\small (2010)}^{\pm }%
\end{array}%
$ & $%
\begin{array}{r}
{\small 1869.7} \\
{\small 2010.3}%
\end{array}%
$ & $%
\begin{array}{r}
{\small 1871} \\
{\small 2010}%
\end{array}%
$ & $%
\begin{array}{r}
{\small D}_{s} \\
{\small D}_{s}^{\ast }{\small [J}^{P}?^{?}{\small ]}%
\end{array}%
$ & $%
\begin{array}{r}
{\small 1968.3} \\
{\small 2112.2}%
\end{array}%
$ & $
\begin{array}{r}
{\small 1969} \\
{\small 2111}%
\end{array}%
$ \\ $%
\begin{array}{rr}
{\small 1}^{3}{\small P}_{0} & {\small 0}^{+} \\
{\small 1P}_{1} & {\small 1}^{+} \\
{\small 1P}_{1} & {\small 1}^{+} \\
{\small 1}^{3}{\small P}_{2} & {\small 2}^{+}%
\end{array}%
$ & $%
\begin{array}{r}
{\small D}_{0}^{\ast }{\small (2400)}^{\pm } \\
{\small D}_{1}{\small (2430)}^{0} \\
{\small D}_{1}{\small (2420)}^{\pm }{\small [J}^{P}?^{?}{\small ]} \\
{\small D}_{2}^{\ast }{\small (2460)}^{\pm }%
\end{array}%
$ & $%
\begin{array}{r}
{\small 2351(7)} \\
{\small 2427(40)} \\
{\small 2423.2} \\
{\small 2465.4}%
\end{array}%
$ & $%
\begin{array}{r}
{\small 2406} \\
{\small 2469} \\
{\small 2426} \\
{\small 2460}%
\end{array}%
$ & $%
\begin{array}{r}
{\small D}_{s0}^{\ast }{\small (2317)} \\
{\small D}_{s1}{\small (2460)} \\
{\small D}_{s1}{\small (2536)} \\
{\small D}_{s2}^{\ast }{\small (2573)}%
\end{array}%
$ & $%
\begin{array}{r}
{\small 2317.7} \\
{\small 2459.5} \\
{\small 2535.1} \\
{\small 2569.1}%
\end{array}%
$ & $%
\begin{array}{r}
{\small 2509} \\
{\small 2574} \\
{\small 2536} \\
{\small 2571}%
\end{array}%
$ \\ $%

\begin{array}{rc}
{\small 1}^{3}{\small D}_{1} & {\small 1}^{-} \\
{\small 1D}_{2} & {\small 2}^{-} \\
{\small 1D}_{2} & {\small 2}^{-} \\
{\small 1}^{3}{\small D}_{3} & {\small 3}^{-}%
\end{array}%
$ & $%
\begin{array}{c}
\\
\\
{\small D(2740)}^{0}{\small [J}^{P}?^{?}{\small ]} \\
{\small D}_{3}^{\ast }{\small (2750)}%
\end{array}%
$ & $%
\begin{array}{c}
\\
\\
{\small 2737(12)} \\
{\small 2763.5}%
\end{array}%
$ & $%
\begin{array}{c}
{\small 2788} \\
{\small 2850} \\
{\small 2806} \\
{\small 2863}%
\end{array}%
$ & $%
\begin{array}{c}
{\small D}_{s1}^{\ast }{\small (2860)} \\
\\
\\
{\small D}_{s3}^{\ast }{\small (2860)}%
\end{array}%
$ & $%
\begin{array}{c}
{\small 2859(27)} \\
\\
\\
{\small 2860(7)}%
\end{array}%
$ & $%
\begin{array}{c}
{\small 2913} \\
{\small 2961} \\
{\small 2931} \\
{\small 2971}%
\end{array}%

$ \\
\hline\hline
\end{tabular}
\end{table*}

\medskip

\renewcommand\tabcolsep{0.47cm}
\renewcommand{\arraystretch}{1.5}
\begin{table*}[!htbp]
\caption{The observed masses (in MeV) of the bottomed and bottomed strange mesons \cite{Tanabashi:D18}. The some of quantum numbers as shown are quark model
predictions. The errors less than $5$ MeV are not indicated.}\label{bm}
\begin{tabular}{ccccccc}
\hline\hline
{\small State }$J^{P}$ & {\small Meson} & {\small Mass} &
{\small EFG \cite{EFG:C10}} & {\small Meson} & {\small Mass} &
{\small EFG \cite{EFG:C10}} \\ \hline
$%
\begin{array}{rr}
{\small 1}^{1}{\small S}_{0} & {\small 0}^{-} \\
{\small 1}^{3}{\small S}_{1} & {\small 1}^{-}%
\end{array}%
$ & $%
\begin{array}{r}
B^{0} \\
B^{\ast }%
\end{array}%
$ & $%
\begin{array}{r}
{\small 5279.6} \\
{\small 5324.7}%
\end{array}%
$ & $
\begin{array}{r}
{\small 5280} \\
{\small 5326}%
\end{array}%
$ & $%
\begin{array}{r}
{\small B}_{s} \\
{\small B}_{s}^{\ast }%
\end{array}%
$ & $%
\begin{array}{r}
{\small 5366.9} \\
{\small 5415.4}%
\end{array}%
$ & $
\begin{array}{r}
{\small 5372} \\
{\small 5414}%
\end{array}%
$ \\ $%
\begin{array}{rr}
{\small 1}^{3}{\small P}_{0} & {\small 0}^{+} \\
{\small 1P}_{1} & {\small 1}^{+} \\
{\small 1P}_{1} & {\small 1}^{+} \\
{\small 1}^{3}{\small P}_{2} & {\small 2}^{+}%
\end{array}%
$ & $%
\begin{array}{r}
{\small B}_{J}^{\ast }{\small (5732)}{\small [J}^{P}?^{?}{\small ]} \\
\\
{\small B}_{1}{\small (5721)}^{0} \\
{\small B}_{2}^{\ast }{\small (5747)}^{0}%
\end{array}%
$ & $%
\begin{array}{r}
{\small 5698(8)} \\
\\
{\small 5726.0} \\
{\small 5739.5}%
\end{array}%
$ & $
\begin{array}{r}
{\small 5749} \\
{\small 5774} \\
{\small 5723} \\
{\small 5741}%
\end{array}%
$ & $%
\begin{array}{r}
\\
{\small B}_{sJ}^{\ast }{\small (5850)}{\small [J}^{P}?^{?}{\small ]} \\
{\small B}_{s1}{\small (5830)} \\
{\small B}_{s2}^{\ast }{\small (5840)}%
\end{array}%
$ & $%
\begin{array}{r}
\\
{\small 5853(15)} \\
{\small 5828.6} \\
{\small 5839.9}%
\end{array}%
$ & $
\begin{array}{r}
{\small 5833} \\
{\small 5865} \\
{\small 5831} \\
{\small 5842}%
\end{array}%

$ \\
\hline\hline
\end{tabular}
\end{table*}

\medskip

\renewcommand\tabcolsep{0.38cm}
\renewcommand{\arraystretch}{1.8}
\begin{table*}[!htbp]
\caption{The effective masses(in GeV) of quarks that match the observed spin-averaged masses in Table VI and VII, with $a$ in GeV and the RMS error $\chi_{RMS}=0.001$GeV. The comparison with that by quark model is given. }\label{em}
\begin{tabular}{cccccccccc}
\hline\hline
{\small Parameters} & ${\small M}_{c}$ & ${\small M}_{b}$ & ${\small m}_{n}$
& ${\small m}_{s}$ & ${\small a}${\small (}${\small c\bar{n}}${\small )} & $%
{\small a}${\small (}${\small c\bar{s}}${\small )} & ${\small a(b\bar{n})}$
& ${\small a(b\bar{s})}$ \\ \hline
{\small This work} & ${\small 1.44[\text{input}]}$ & ${\small 4.48[\text{input}]}$ & ${\small 0.23}$ & $%
{\small 0.328}$ & ${\small 0.223}$ & ${\small 0.249}$ & ${\small 0.275}$ & $%
{\small 0.313}$ \\
{\small EFG \cite{EFG:C10}} & ${\small 1.55}$ & ${\small 4.88}$ & ${\small 0.33}$ & $%
{\small 0.5}$ & ${\small 0.64}${\small /}${\small 0.58}$ & ${\small 0.68/0.64%
}$ & ${\small 1.25/1.21}$ & ${\small 1.28/1.23}$ \\
\hline\hline
\end{tabular}
\end{table*}

\renewcommand\tabcolsep{0.8cm}
\renewcommand{\arraystretch}{1.5}
\begin{table*}[!htbp]
\caption{The measured spin-averaged masses of the SH baryons and that predicted by the linear Regge relation (\ref{GRegg}), with the parameters in Tables III-V. }\label{em}
\begin{tabular}{c|cc|ccc}
\hline\hline
Baryon &  Exp $(\bar{M}) $&This work $(\bar{M})$ & Baryon &  Exp $(\bar{M}) $ &  This work $(\bar{M})$
\\ \hline
$\Lambda_{c},1S$ &  $2286.0 $& $2286.0$ & $\Xi_{c},1S$ & $2467.87 $ & $ 2469.0$
\\
$\Lambda_{c},1P$ &  $2616 $& $2615.5$ & $\Xi_{c},1P $& $2808 $ &  $2803.9$
\\
$\Lambda_{c},1D$ &  $2871 $& $2871.1$ & $\Xi_{c},1D $& $3069 $ &  $3071.3$
\\ \hline
$\Lambda_{b},1S$ &  $5619 $& $5619$ & $\Xi_{c},1S $& $5792 $ &  $5798.1$
\\
$\Lambda_{b},1P$ &  $5917 $& $5616.3$ & $\Xi_{c},1P $& $-- $ &  $6125.1$
\\
$\Lambda_{b},1D$ &  $6148 $& $6149.3$ & $\Xi_{c},1D $& $-- $ &  $6380.9$
\\ \hline
$\Sigma_{c},1S$ &  $2497.0 $& $2496.1$ & $\Xi_{c}^{'},1S $& $2624 $ &  $2623.1$
\\
$\Sigma_{c},1P$ &  $-- $& $2774.1$ & $\Xi_{c}^{'},1P $& $-- $ &  $2923.0$
\\ \hline
$\Sigma_{b},1S$ &  $5825 $& $5825.0$ & $\Xi_{b}^{'},1S $& $5949 $ &  $5949.0$
\\
$\Sigma_{b},1P$ &  $--$& $6088.4$ & $\Xi_{b}^{'},1P $& $--$ &  $6248.2$
\\

\hline\hline
\end{tabular}
\end{table*}
In Table III, IV and V, the trajectory parameters reproducing the observed masses in Table I and Table II are listed for the charmed baryons $\Lambda _{c}$ and $\Xi _{c}$, and for the baryons $\Sigma _{c}$ and charm-strange baryons $\Xi _{c}^{\prime }$, respectively, including the ensuing $P$-wave spin-averaged masses predicted by Eq.(\ref{MnL}). The resulted parameters are compared with the quark model predictions \cite{EFG:D11} in the three tables. In table IX, the observed spin-averaged masses are collected for all relevant baryons except for the baryons $\Sigma _{Q}$ and $\Xi _{Q}^{\prime}$ ($Q=c,b$), and the spin-averaged masses predicted by Eq.(\ref{GRegg}) via the parameters given in Table III,IV and V are also shown for comparisons. Here, we assumed the orbital trajectories of the $\Sigma_{c}$($\Xi_{c}$) and the $\Lambda_{c}$($\Xi_{c}^{\prime }$) to be parallel, leading to the same string tensions between them.

We also apply Eq.(\ref{GRegg}), with $m_{d}$ everywhere replaced by the effective masses of light(strange or $u/d$) antiquarks, to the heavy-light ($Q\bar{q}$) mesons whose observed masses are listed in Tables VI and VII, and list in Table VIII the determined trajectory parameters, namely, the effective masses $m_{n,s}$ of light $u/d$ quark and the strange quark, and the string tensions $a$,  for the $D/D_{s}$ mesons and the $B/B_{s}$ mesons. The resulted parameters are compared to that in the quark model \cite{EFG:C10} for the heavy-light mesons.

From Table III one sees that the predicted spin-averaged masses $2618$\ \text{MeV} and $2804$\ \text{MeV} of the $1P$-wave $\Lambda _{c}$ and $\Xi _{c}$ are very close to the corresponding values $2616$\ MeV and $2808$\ MeV in Tables I, respectively. Furthermore, the following remarks are in order.

(1) The mass differences between the S-wave SH baryons $\Sigma_{c}$ and $\Lambda _{c}$ are equal to that between the diquarks (of scalar or vector type) contained in them,
\begin{equation}
m_{d}(\Sigma_{c})-m_{d}(\Lambda _{c})=210\text{MeV}=\bar{M}(\Sigma_{c})-M(\Lambda _{c}), \label{Cdeg}
\end{equation}
\begin{equation}
m_{d}(\Xi _{c}^{\prime })-m_{d}(\Xi_{c})=154\text{MeV}=\bar{M}(\Xi _{c}^{\prime })-M(\Xi_{c}).
\label{Xdeg}
\end{equation}
This imposes a constraint on the masses of the vector diquark which are crucial for predicting, through Eq.(\ref{GRegg}), the spin-averaged masses of the $P$-wave excitations of the baryons $\Sigma _{c}$ and $\Xi _{c}^{\prime }$. The latter are less established experimentally, as shown in Table II.

(2) The mass values of the heavy quark and light diquarks in our approach are in consistent with that in the relativistic quark model. The string tension $a$ in our approach is flavor-dependent weakly, with a rising about $0.04$GeV$^{2}$ ($0.06$GeV$^{2}$) for the charmed(bottom) baryons when diquark flavor in them changes from $nn$ to $ns$($n=u,d$).

(3) Once determined, the diquark masses of both types can be applied to explore the Regge behaviors in the bottom partner of the charmed baryons, for which the less data are available experimentally.

\section{Including mass splitting for orbital excitations}

To estimate the mass splitting in orbitally excited charmed and bottom
baryons $Qqq$($Q=c,b,q=u/d,s, qq\neq ss$), one needs to consider the spin-dependent
interaction which was ignored in Eq.(\ref{GRegg}). In the light of QCD
string, the orbital motion of QCD string[34] contributes dominantly to total
angular moment $L$ of the system, which makes the quark-diquark potential
nontrivial($L$-dependent) effectively. For simplicity, we follow Ref. \cite{KarlinerR:D15} to compute the
spin-dependent mass for $L=0,1$ and $2$($\Lambda _{Q},\Xi _{Q}$ case only)
via the scaling relation between heavy baryons $Qqq$ and their meson partners $Q%
\bar{q}$(note that the color-structure similarity between baryon and meson
is implicitly assumed in the quark-diquark picture). The spin-dependent
interaction we consider is\cite{EFG:D11,KarlinerR:D15}
\begin{equation}
H^{SD}=a_{1}\mathbf{L}\cdot \mathbf{S}_{d}+a_{2}\mathbf{L}\cdot \mathbf{S}%
_{Q}+b\mathbf{S}_{12}+c\mathbf{S}_{d}\cdot \mathbf{S}_{Q},  \label{Hsd}
\end{equation}%
\begin{equation*}
\mathbf{S}_{12}=3\mathbf{S}_{d}\cdot \mathbf{\hat{r}S}_{Q}\cdot \mathbf{\hat{%
r}}-\mathbf{S}_{d}\cdot \mathbf{S}_{Q}\mathbf{,}
\end{equation*}%
where the first two terms are spin-orbit energies, the third is the tensor
energy, and the last describes the hyperfine splitting between the heavy
quark with spin $\mathbf{S}_{Q}$ and diquark with spin $\mathbf{S}_{d}$.
Here, $\mathbf{L}$ is the orbital angular momentum of the baryons. The first
term($a_{1}$-term) is obviously larger than the others (especially for the
bottom systems) as these terms all involve the magnetic moment $\mathbf{S}%
_{Q}/M_{Q}$ of heavy quark when viewing Eq.(\ref{Hsd}) as a spin-relevant
relativistic correction.

(A) The S-wave baryons($L=0$). In this case, only the last term survives in
Eq.(\ref{Hsd}), $H^{SD}=c\mathbf{S}_{d}\cdot \mathbf{S}_{Q}$. Note that $%
\left\langle \mathbf{S}_{d}\cdot \mathbf{S}_{Q}\right\rangle
=[S(S+1)-S_{Q}(S_{Q}+1)-S_{d}(S_{d}+1)]/2$ becomes $0$ when the conserved
total quark spin $S$($=J$ here)$=1/2$ and $S_{d}=0$, and has the value $-1$
and $1/2$, when $S_{d}=1$ and $S=1/2,3/2$, respectively. Thus, the mass for
the $1S$-wave states is
\begin{eqnarray}
M(1S)_{S_{d}=0} &=&\bar{M}_{L=0}\text{, (}\Lambda_{Q}\text{/}\Xi _{Q}%
\text{),}  \label{M1S1} \\
M(1S)_{S_{d}=1} &=&\bar{M}_{L=0}+c\left[
\begin{array}{rr}
-1 & 0 \\
0 & 1/2%
\end{array}%
\right] ,\text{(}\Sigma _{Q}\text{/}\Xi _{Q}^{\prime }\text{).}
\label{M1S2}
\end{eqnarray}

For the baryon $\Lambda _{Q}$ and $\Xi _{Q}$ containing an light scalar
diquark(spin $S_{d}=0$), the total spin $S=1/2=J$. This corresponds to a
single baryon ground state $1^{2}S_{1/2}$. The ground state masses of the
baryons $\Lambda _{Q}$ and $\Xi _{Q}$ are given simply by their
spin-independent masses in Eq.(\ref{GRegg}). To estimate $c$ for the baryon
$\Sigma _{Q}$ and $\Xi _{Q}^{\prime }$ which contain the light vector
diquark(spin $S_{d}=1$), we employ the following scaling relation to the partner in heavy mesons,%
\begin{equation}
c(Qqq)=\frac{m_{q}}{m_{d}(qq)}c(Q\bar{q}),  \label{Sc}
\end{equation}%
which stems from the inverse-mass scaling inherent in the contact
interaction ($\sim $ the diquark magnetic moment $\mathbf{S}_{d}/m_{d}$).
Extracting $c(Q\bar{q})$ from the splitting between the $Q\bar{q}$ meson
mass $M(1^{3}S_{1},1^{-})$ and $M(1^{1}S_{0},0^{-})$ in the case of the $D$ mesons(see Table VI):%
\begin{equation}
c(Q\bar{q})=M(1^{3}S_{1},1^{-})-M(1^{1}S_{0},0^{-})=140.6\text{MeV},
\label{cM}
\end{equation}%
one estimates, by Eq.(\ref{Sc}),%
\begin{equation}
c=140.6 \times \left\{
\begin{array}{r}
230/745 \\
230/872%
\end{array}%
\right\} =\allowbreak \left\{
\begin{array}{r}
43.4\text{(}\Sigma _{c}\text{)} \\
37.1\text{(}\Xi _{c}^{\prime }\text{)}%
\end{array}%
\right\} \text{,}  \label{cMN}
\end{equation}%
Having $c$ given by Eps.(\ref{cMN}), Eqs.(\ref{M1S1}-\ref{M1S2}) give
masses for the S-wave charmed baryons, as listed in Table I and Table II,
respectively.

(B) The P-wave baryons($L=1$). For the baryon $\Lambda _{Q}$ and $\Xi _{Q}$
containing spin$=0$ diquarks, the P-wave states correspond to two baryon
states of $J^{P}=1/2^{-}$ and $3/2^{-}$. This can be written as $\left[
0_{d}\otimes \left( \frac{1}{2}\right) _{Q}\right] _{S}\otimes 1_{L}=$ $%
\frac{1}{2}\oplus \frac{3}{2}$ in the notation of symmetry group, where the
notations $\otimes $ and $\oplus \,$ stand for the coupling(adding of the
angular momentum in the standard terminology) and the juxtaposition of the
angular momentum shown. In the charm sectors, for instance, there are four
P-wave candidates $\Lambda _{c}(2595)^{+}\ $and $\Lambda _{c}(2625)^{+}$, $%
\Xi _{c}(2790)^{+}$ and $\Xi _{c}(2815)^{+}$\cite{Tanabashi:D18} (see Table
I). The masses and possible quantum numbers for the observed low-lying
baryons $\Lambda _{c}$ and $\Xi _{c}$ are summarized in Table I.

For the baryon $\Sigma _{Q}$ and $\Xi _{Q}^{\prime }$ containing spin=1 diquarks, the five P-wave states $J=1/2^{^{\prime }}$,$3/2^{\prime }$,$1/2$,,%
$3/2$ and $5/2$ with negative parity are allowed:%
\begin{eqnarray}
\left[ 1_{d}\otimes \left( \frac{1}{2}\right) _{Q}\right] _{S}\otimes 1_{L}
&=&\left[ \frac{1}{2}^{\prime }\oplus \frac{3}{2}\right] _{S}\otimes 1_{L},
\notag \\
&=&\frac{1}{2}\oplus \frac{3}{2}\oplus \frac{1}{2}^{\prime }\oplus \frac{3}{2%
}^{\prime }\oplus \frac{5}{2}.  \label{LS}
\end{eqnarray}%
The current candidates of these $P$-wave charmed baryons are the $\Sigma
_{c}(2800)^{0,+,++}$ in the nonstrange sector and the $\Xi _{c}(2930)^{0,+}$ in
the strange sector, whose spin and parity have not been established yet\cite%
{Tanabashi:D18}. The quark model assignments are given by Eq.(\ref{JP}). The
masses and possible quantum numbers for the observed low-lying baryons $%
\Sigma _{c}$ and $\Xi _{c}^{\prime }$ are summarized in Table II, with the
question marks indicating that their $J^{P}$ assignments are unknown experimentally.

A normal scheme(the $L$-$S$ coupling) of computing mass splitting by Eq. (%
\ref{Hsd}) is to couple the diquark spin $\mathbf{S}_{d}$ to the heavy quark
spin $\mathbf{S}_{Q}$ first and then to the orbital angular momentum $%
\mathbf{L}$ of the system, as shown in (\ref{LS}) for $S_{d}=1$ case. In the $L$-$S$ basis the two $J=1/2$ states and the two $J=3/2$ states are usually mixed
unless $a_{1}=a_{2}$. They are eigenstates of $2\times 2$ matrices $M_{J}$
given in the basis \{$^{2}P_{J},^{4}P_{J}$\}. Using the $1P$-wave
wavefunctions of the heavy quark-diquark systems in the basis of $%
|L_{z},S_{Qz},S_{dz}\rangle $, these matrices can be evaluated and given by
(Appendix B)
\begin{equation*}
\left\langle \mathbf{L}\cdot \mathbf{S}_{d}\right\rangle _{J=\frac{1}{2}}=%
\left[
\begin{array}{rr}
-\frac{4}{3} & -\frac{\sqrt{2}}{3} \\
-\frac{\sqrt{2}}{3} & -\frac{5}{3}%
\end{array}%
\right]
\end{equation*}%
\begin{equation*}
\left\langle \mathbf{L}\cdot \mathbf{S}_{d}\right\rangle _{J=\frac{3}{2}}=%
\left[
\begin{array}{rr}
\frac{2}{3} & -\frac{\sqrt{5}}{3} \\
-\frac{\sqrt{5}}{3} & -\frac{2}{3}%
\end{array}%
\right]
\end{equation*}%
and $\left\langle \mathbf{L}\cdot \mathbf{S}_{d}\right\rangle _{J=5/2}=1$, $%
\left\langle \mathbf{L}\cdot \mathbf{S}_{Q}\right\rangle _{J=5/2}=1/2$. The
two matrices above have the eigenvalues $-1$, $-2$ for $J=1/2$ and the
eigenvalues $\pm 1$ for $J=3/2$.

One can also choose an alternative scheme(the $j$-$j$ coupling) in which
the diquark spin $\mathbf{S}_{d}$ couples to $\mathbf{L}$ first and then to
the conserved spin $\mathbf{S}_{Q}$ of the heavy quark. In the P-wave case($%
L=1$), for instance, this scheme reads,%
\begin{equation*}
\left[ 0_{d}\otimes 1_{L}\right] _{j}\otimes \left( \frac{1}{2}\right)
_{Q}=1_{j}\otimes \left( \frac{1}{2}\right) _{Q}=\frac{1}{2}\oplus \frac{3}{2%
},
\end{equation*}%
for the baryons $\Lambda _{Q}$ and $\Xi _{Q}$, and takes form%
\begin{eqnarray*}
\left[ 1_{d}\otimes 1_{L}\right] _{j}\otimes \left( \frac{1}{2}\right) _{Q}
&=&\{0,1,2\}_{j}\otimes \left( \frac{1}{2}\right) _{Q} \\
&=&\frac{1}{2}\oplus \frac{1}{2}^{\prime }\oplus \frac{3}{2}\oplus \frac{3}{2%
}^{\prime }\oplus \frac{5}{2},
\end{eqnarray*}%
for the baryons $\Sigma _{Q}$ and $\Xi _{Q}^{\prime }$. Since the interactions other than the $\mathbf{L}\cdot \mathbf{S}_{d}$ term in (\ref{Hsd}) are suppressed by $1/M_{Q}$, one can compute the mass
splitting in a basis in which $\mathbf{L}\cdot \mathbf{S}_{d}$ is diagonal,
with the other interactions treated perturbatively.

Notice that $\left\langle \mathbf{L}\cdot \mathbf{S}_{d}\right\rangle
=[j(j+1)-L(L+1)-S_{d}(S_{d}+1)]/2$ takes values $-2$, $-1$ and $1$ when $j=0$%
, $1,2$, respectively, one can use the eigenfunction of the $\mathbf{L}\cdot
\mathbf{S}_{d}$ to evaluate the mass splitting operators in Eq.(\ref{Hsd})
in the $j$-$j$ coupling(Appendix B). The results are listed in Table X,
from which the mass splitting $M(J,j)=\langle J,j|H^{SD}|J,j\rangle $ can
be evaluated in the $|J,j\rangle $ basis. Written in terms of three
parameters ($a_{1},a_{2}+c,b$), the results are(see Ref. \cite{KarlinerR:D2017}) for details, which corrected an error in Ref. \cite{KarlinerR:D15})

\renewcommand\tabcolsep{0.6cm}
\renewcommand{\arraystretch}{1.5}
\begin{table}
\caption{The matrix elements of the mass splitting operators in the P-wave
baryon states in the $j$-$j$ coupling }\label{em}
\begin{tabular}{cccc}
\hline\hline
($J,j$) & $\left\langle \mathbf{L}\cdot \mathbf{S}_{Q}\right\rangle $ & $%
\left\langle \mathbf{S}_{12}\right\rangle $ & $\left\langle \mathbf{S}%
_{d}\cdot \mathbf{S}_{Q}\right\rangle $ \\ \hline
($1/2,0$) & $0$ & $0$ & $0$ \\
($1/2,1$) & $-\frac{1}{2}$ & $-1$ & $-\frac{1}{2}$ \\ \hline
($3/2,1$) & $\frac{1}{4}$ & $\frac{1}{2}$ & $\frac{1}{4}$ \\
($3/2,2$) & $-\frac{3}{4}$ & $\frac{3}{10}$ & $-\frac{3}{4}$ \\ \hline
($5/2,2$) & $\frac{1}{2}$ & $-\frac{1}{5}$ & $\frac{1}{2}$ \\ \hline\hline
\end{tabular}
\end{table}

\begin{eqnarray}
M^{SD}(1/2,0) &=&-2a_{1},  \label{hsd1} \\
M^{SD}(1/2,1) &=&-a_{1}-\frac{1}{2}(a_{2}+c)-b,  \label{hsd2} \\
M^{SD}(3/2,1) &=&-a_{1}+\frac{1}{4}(a_{2}+c)+\frac{1}{2}b,  \label{hsd3} \\
M^{SD}(3/2,2) &=&a_{1}-\frac{3}{4}(a_{2}+c)+\frac{3}{10}b,  \label{hsd4} \\
M^{SD}(5/2,2) &=&a_{1}+\frac{1}{2}(a_{2}+c)-\frac{1}{5}b.  \label{hsd5}
\end{eqnarray}%
One additional constraint stems from the vanishing of these splitting as a
sum:$\sum (2J+1)M^{SD}(J,j)=0$, which is
\begin{equation}
10M^{SD}(1/2,0)-15M^{SD}(1/2,1)+8M^{SD}(3/2,2)=3M^{SD}(5/2,2).  \label{SM}
\end{equation}

Following Ref. \cite{KarlinerR:D15}, one can calculate the parameters $%
\{a_{1},a_{2},b\}$ in Eq.(\ref{Hsd}) for the heavy baryon $Qqq$($q=u/d$, $s$%
) by assuming the following mass scaling, which relates these parameters to
that for the heavy mesons $Q\bar{q}$ with effective antiquark mass $m_{q}$,
or relates the parameters among the heavy flavor partners($bqq$ and $cqq$%
),
\begin{equation}
a_{1}\text{(}Qqq\text{)}=a_{1}(Q\bar{q})\left( \frac{m_{q}}{m_{d}\text{(}qq%
\text{)}}\right) \text{,}  \label{scaling}
\end{equation}%
\begin{equation}
a_{1}\text{(}b\text{)=}a_{1}\text{(}c\text{),}a_{2}\text{(}b\text{)=}a_{2}%
\text{(}c\text{)}\left( \frac{M_{c}}{M_{b}}\right) \text{, }b\text{(}b\text{%
)=}b\text{(}c\text{)}\left( \frac{M_{c}}{M_{b}}\right) \text{,}
\label{scal2}
\end{equation}%
where $b,c$ in parenthesis stand for the bottom and charm sectors. In
evaluating the coupling $a_{2}$ for the charmed baryons, a simple way is to
estimate it via that for the charmed mesons by resembling the charmed
baryon $cqq$ to the charmed meson $c\bar{q}$ systems. Considering the
finite mass effects of the charm quark which differ between the charmed
baryons and the charmed mesons, we use a recoil factor $(1+m_{d}/M_{Q})$ to
partially remedy the uncertainty that may happen in this resembling,
\begin{equation}
a_{2}(cqq)=\frac{a_{2}(c\bar{q})}{1+m_{d}/M_{c}}.  \label{res}
\end{equation}

In the following, we describe very briefly how to evaluate the parameters $%
\{a_{1},a_{2},b\}$ for the charmed-strange mesons for which the
spin-dependent Hamiltonian is given by Eq.(\ref{Hsd}) in which $\mathbf{S}_{d}$ is replaced by $\mathbf{S}_{q}$. Firstly, one can
derive a counterpart of Eqs.(\ref{hsd1}-\ref{hsd5}) for $Q\bar{q}$ system
via a calculation similar to the P-wave heavy baryons and obtain the mass
for the P-wave spin multiplets of the heavy mesons($Q\bar{q}$) in the $%
|J,j\rangle $ states($j=L_{3}+S_{3},L=1$). The results are(Appendix C:
Eq.(C8-C11)):
\begin{eqnarray}
M(0,1/2) &=&\bar{M}-a_{1}-a_{2}-b-\frac{c}{4},  \label{C8} \\
M(1,1/2) &=&\bar{M}-a_{1}+\frac{1}{3}a_{2}+\frac{1}{3}b-\frac{c}{12},
\label{C9} \\
M(1,3/2) &=&\bar{M}+\frac{1}{2}a_{1}-\frac{5}{6}a_{2}+\frac{1}{6}b-\frac{5c}{%
12},  \label{C10} \\
M(2,3/2) &=&\bar{M}+\frac{1}{2}a_{1}+\frac{1}{2}a_{2}-\frac{1}{10}b+\frac{c}{%
4}.  \label{C11}
\end{eqnarray}%
Secondly, confronting the observed masses of the $D_{s}$ mesons in the
P-wave(see Table VI), one can find the parameters $\{a_{1},a_{2},b\}$ for
the heavy mesons $D_{s}$ reproducing their observed mass spectrum via Eqs.(%
\ref{hsd1}-\ref{hsd5}). Ignoring $c$(vanishes almost in the P-wave), one can
find lastly\cite{KarlinerR:D15}(see Appendix C: Eq.(C12)):%
\begin{equation}
\begin{Bmatrix}
\bar{M}\text{(MeV),} & a_{1}\text{(MeV),} & a_{2}\text{(MeV)} & b\text{(MeV)}
\\
2513.4, & 89.4, & 40.7, & 65.6.%
\end{Bmatrix}
\label{C12}
\end{equation}

Given $a_{1}$ in Eq.(\ref{C12}) for the $D_{s}$ mesons in P-wave, the
diquark masses in Table III and IV and the strange quark mass in Table VIII,
Eq. (\ref{scaling}) gives%
\begin{equation}
\begin{array}{rr}
\Sigma _{c},1P: & \ a_{1}=a_{1}(c\bar{s})\left( \frac{m_{s}}{%
m_{d}(\{qq\})}\right) =89.4\text{MeV}\left( \frac{328}{745}\right)
=\allowbreak 39.4\text{MeV,} \\
\Xi _{c}^{^{\prime }},1P: & a_{1}=a_{1}(c\bar{s})\left( \frac{m_{s}}{%
m_{d}(\{qs\})}\right) =89.4\text{MeV}\left( \frac{328}{872}\right) =33.6%
\text{MeV,}%
\end{array}
\label{Sgm12}
\end{equation}

For the charmed baryons $\Sigma _{c}$ and $\Xi _{c}^{\prime }$, Eq.(\ref{res}%
) with $m_{d}$ given in Table IV gives $\allowbreak $
\begin{equation}
\begin{array}{r}
a_{2}(\Sigma _{c},1P)=\frac{a_{2}(D_{s})}{1+m_{d}/M_{c}}=\frac{40.7}{1.517}%
=26.8\text{MeV,} \\
a_{2}(\Xi _{c}^{\prime },1P)=\frac{a_{2}(D_{s})}{1+m_{d}/M_{c}}=\frac{40.7}{%
1.606}=25.3\text{MeV,}%
\end{array}
\label{a2charm}
\end{equation}%
while for the charmed baryons $\Lambda _{c}$ and $\Xi _{c}$, the parameter $%
a_{2}$ become slightly larger when using Eq.(\ref{C12}) and the diquark
masses given in Table III. We take
$a_{2}(\Lambda _{c},1P)\simeq a_{2}(\Sigma _{c},1P)$ and  $a_{2}(\Xi
_{c},1P)\simeq a_{2}(\Xi _{c}^{\prime },1P)$ for simplicity as this varying in $a_{2}$ is small at the mass scale of the P-wave charmed baryons, as shown explicitly in FIGs. 1-2.  Eq.(\ref{a2charm}) are in consistent with the values of $a_{2}$ that match the
observed P-wave masses of the baryons $\Lambda _{c}$ and $\Xi _{c}$:
\begin{equation}
a_{2}(\Lambda _{c})=24\text{MeV, }a_{2}(\Xi _{c})=16\text{MeV.}  \label{a2SX}
\end{equation}%
These two values follows from the mass differences (see Table I) between the
$J=3/2^{-}$ and $J=1/2^{-}$ states of the baryon $\Lambda _{c}$ and the $\Xi
_{c}$, respectively, which depend only on the second term in Eq.(\ref{Hsd})
when $\mathbf{S}_{d}=0$, that is, $M(J=3/2^{-})-M(J=1/2^{-})=3a_{2}/2$.

As $\mathbf{S}_{d}=0$ for the $\Lambda _{c}$ and the $\Xi _{c}$, the mass
splitting operator for them simplifies, $H^{SD}=a_{2}\mathbf{L}\cdot \mathbf{%
S}_{Q}$, which gives the mass in the $|J=1/2,j=1\rangle $ and $%
|J=3/2,j=1\rangle $:
\begin{equation}
M(\Lambda _{c}/\Xi _{c},1P)=\bar{M}_{L=1}+a_{2}(\Lambda _{c}/\Xi _{c})\left[
\begin{array}{rr}
-1 & 0 \\
0 & 1/2%
\end{array}%
\right] ,  \label{M1P2}
\end{equation}%
where the expectation values of $\mathbf{L}\cdot \mathbf{S}_{Q}$ is given by
$[J(J+1)-L(L+1)-S_{Q}(S_{Q}+1)]/2$\thinspace , taking values $-1$ for $%
J=1/2\,$and $+1/2$ for $J=3/2$. The computed masses via Eqs. (\ref{M1P2}) and (\ref{a2charm}) for the P-wave $\Lambda _{c}/\Xi _{c}$ are collected in Table I.

Putting all together and ignoring the spin-spin $c$-term(quite small) in the
$P$-wave, it follows from Eqs.(\ref{hsd1}-\ref{hsd5}) that the mass
splitting in the P-wave are
\begin{eqnarray}
M(1/2,0) &=&2694.74,  \label{Msig1} \\
M(1/2,1) &=&2722.64-b,  \label{Msig2} \\
M(3/2,1) &=&2740.37+\frac{1}{2}b,  \label{Msig3} \\
M(3/2,2) &=&2795.63+\frac{3}{10}b,  \label{Msig4} \\
M(5/2,2) &=&2825.63-\frac{1}{5}b.  \label{Msig5}
\end{eqnarray}%
for the baryon $\Sigma _{c}$, and
\begin{eqnarray}
M(1/2,0) &=&2855.76,  \label{Mx1} \\
M(1/2,1) &=&2881.38-b,  \label{Mx2} \\
M(3/2,1) &=&2893.38+\frac{1}{2}b,  \label{Mx3} \\
M(3/2,2) &=&2938.62+\frac{3}{10}b,  \label{mx4} \\
M(5/2,2) &=&2964.62-\frac{1}{5}b,  \label{Mx5}
\end{eqnarray}%
for the baryon $\Xi _{c}^{^{\prime }}$, which are plotted against $b$ in FIG.1 and FIG.2,respectively. Each pair of the adjacent lines corresponds to the choice $a_{2}$ given in Eq. (\ref{a2charm}) (dotted lines) and Eq. (\ref{a2SX}) (solid lines), respectively. As shown in FIG. 1, for moderate values of $b$, there is a clear separation between the three lowest masses $M(1/2,j=0)$,$M(1/2,j=1)$,$M(3/2,j=1)$, and the two highest masses $M(3/2,j=2)$ and $M(5/2,j=2)$.

Without loss of the generality, we assume that $b$ is of same order with $%
a_{2}$(note that $a_{2}=b$ in the quark model\cite{CahnJack:D03} of the
heavy-light mesons in the heavy quark limit). The mass function $M(3/2,j=2)$
in FIG.1 approaches the observed mass of the $\Sigma
_{c}(2800)^{0,+,++}$ at $b$ around 20MeV, while the mass function $M(3/2,j=2)$ meets the observed averaged mass 2936MeV of the $\Xi_{c}(2930)$ and $\Xi_{c}(2942)$, Also, the mass function $M(3/2,j=1)$ slowly tends to
approach the mass of the $\Xi_{c}(2930)$ at larger $b$ in FIG. 2. In both cases, the P-wave states with the spin-parity $J^{P}=3/2^{-}$ is favored. The states with $J^{P}=3/2^{-}$ and $5/2^{-}$ can be the states most likely being relatively narrow and thereby easier to observe when they decay via the higher partial wave(D-wave), as discussed in section IV.

(C) The D-wave baryons($L=2$). We consider only the baryon $\Lambda _{Q}$
and $\Xi _{Q}$ containing spin$=0$ diquarks due to the limited
experimental information for the charmed mesons($c\bar{q}$) in D-wave on which our calculation depends. Two possible D-wave states are the baryon states of $J^{P}=3/2^{+}
$ and $5/2^{+}$, which can be written as
\begin{equation*}
\left[ 0_{d}\otimes \left(\frac{1}{2}\right) _{Q}\right] _{S}\otimes 2_{L}=%
\frac{3}{2}\oplus \frac{5}{2}.
\end{equation*}%
In the charm sectors, for instance, there are four D-wave candidates, the
$\Lambda _{c}(2860)^{+}$, the $\Lambda _{c}(2880)^{+}$, the $\Lambda
_{c}(3055)^{+}$ and the $\Xi _{c}(3080)^{+}$\cite{Tanabashi:D18}, as shown in Table I where the other masses and their quantum numbers are also summarized.

Given $\mathbf{S}_{d}=0$ for the baryon $\Lambda _{Q}$ and $\Xi _{Q}$ the
mass splitting operator is $H^{SD}=a_{2}\mathbf{L}\cdot \mathbf{S}_{Q}$,
which gives the mass in the basis $|3/2,j=2\rangle $ and $|5/2,j=2\rangle $:
\begin{equation}
M(\Lambda _{Q}/\Xi _{Q},1D)=\bar{M}_{L=2}+a_{2}\left[
\begin{array}{rr}
-3/2 & 0 \\
0 & 1%
\end{array}%
\right] .  \label{HDw}
\end{equation}
Employing the parameterized mass in Eqs.(C19-C22) stemming from Eq.(%
\ref{Hsd}) in the $j$-$j$ coupling and confronting it with the observed
masses of the $D$ mesons in D-wave in Table XII(see Appendix C), one finds $a_{2}=12.6$MeV(see  Eq.(C23) for details). It follows from Eq.(\ref{res}), with
$m_{d}$ in Table III,
\begin{equation}
a_{2}\left( cqq\right) =\frac{12.6\text{MeV}}{1+m_{d}/1.44}=\left\{
\begin{array}{r}
9.2\text{MeV(}\Lambda _{c}\text{)} \\
8.4\text{MeV(}{\small \Xi }_{c}\text{)}%
\end{array}%
\right\} .  \label{a2D}
\end{equation}%
by which all computed masses of the D-wave charmed baryons in Table I follow.

\begin{figure}[!h]
\centering
\begin{tabular}{c}
\includegraphics[width=0.44\textwidth]{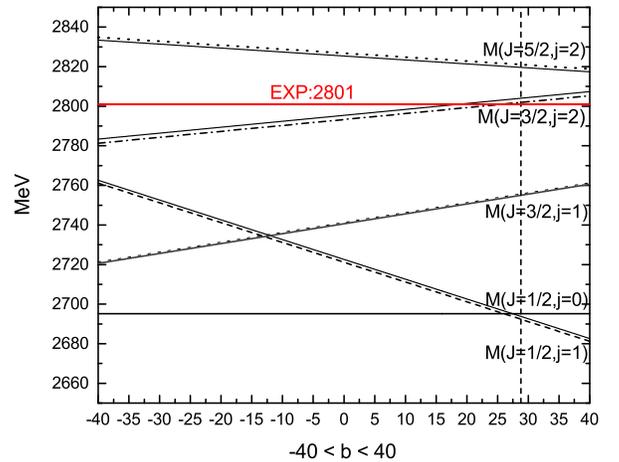}
\end{tabular}
\caption{The masses of the $P$-wave $\Sigma _{c}$ against the parameter $b$.}\label{potentialshape}
\end{figure}
\begin{figure}[!htbp]
\centering
\begin{tabular}{c}
\includegraphics[width=0.44\textwidth]{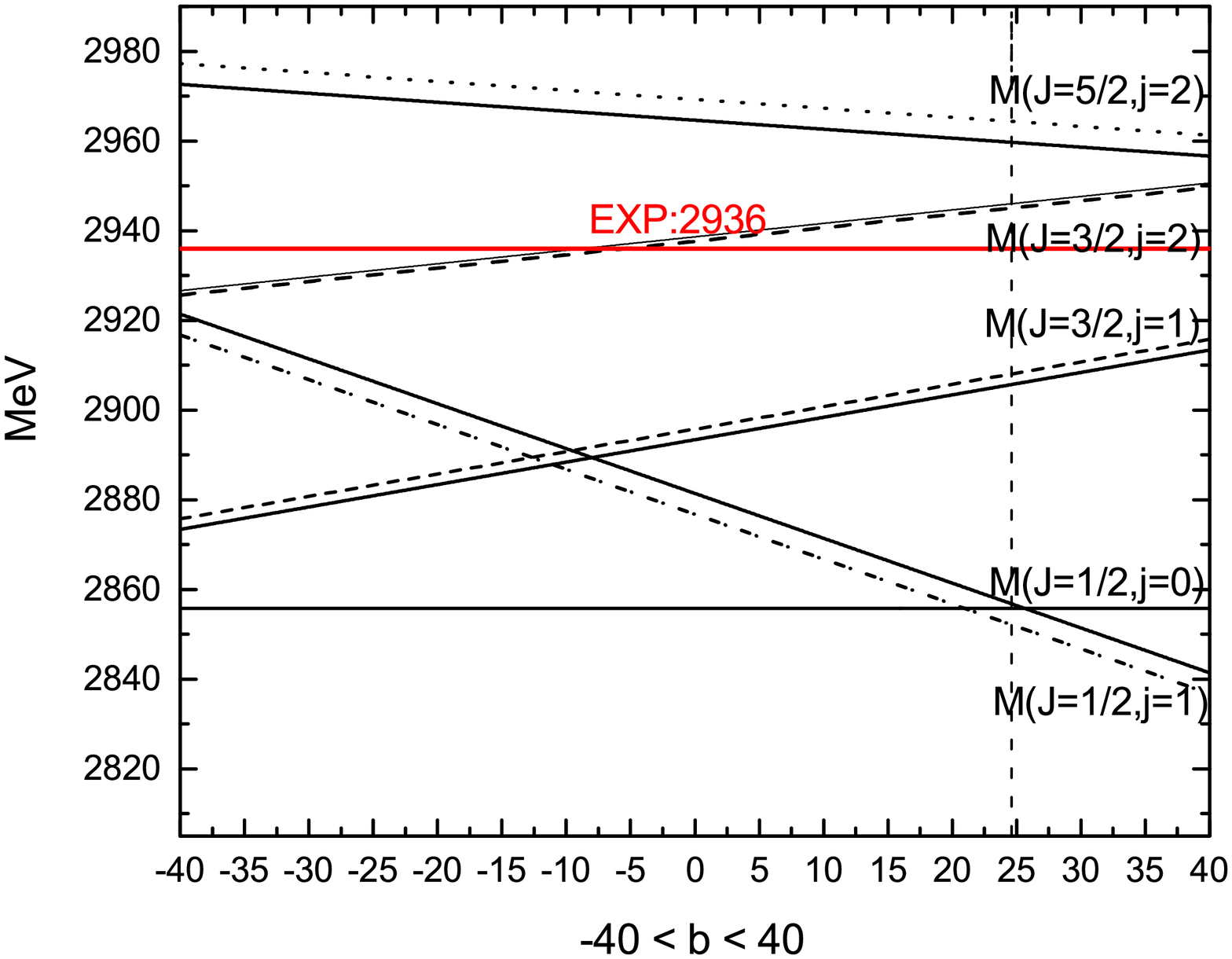}
\end{tabular}
\caption{The masses of the $P$-wave $\Xi _{c}^{\prime }$ against the parameter $b$.}\label{potentialshape}
\end{figure}
\medskip
\section{Mass splitting for the singly bottom baryons}

One can apply the calculational method similar to section III to the
singly bottom baryons. Differing from the charmed baryons, one can improve
the computational accuracy for the bottom baryons via the scaling relation to
the charmed baryons(note that there are few higher partial waves for the $%
B_{s}$ mesons in Table VII). The masses and the quantum numbers for the
observed low-lying baryons $\Lambda _{b}$ and $\Xi _{b}$ are summarized in
Table I, whereas in Table II, we list mainly the observed lowest baryons $%
\Sigma _{b}$ and $\Xi _{b}^{\prime }$ and the newly-observed excited baryons
$\Sigma _{b}(6097)$ and $\Xi _{b}^{^{\prime }}(6227)$, including the
resonances $\Lambda _{b}(6146)$ and $\Lambda_{b}(6152)$\cite{Lamdab19}
discovered very recently by the LHCb Collaboration.

(A) The S-wave baryons($L=0$). Only the last(contact) term in Eq. (\ref{Hsd}%
) survives for the bottom baryons($\Sigma _{b},\Xi _{b}^{\prime }$) while
Eq. (\ref{Hsd}) vanishes for the bottom baryons($\Lambda _{b},\Xi _{b}$) for
which both $\mathbf{L}$ and $\mathbf{S}_{d}$ vanish. Considering the contact
term is suppressed by the magnetic moment $\mathbf{S}_{Q}/M_{Q}$, the
computation of the parameter $c$ for the bottom baryons($\Sigma _{b},\Xi
_{b}^{\prime }$) can be obtained by the following heavy-quark mass scaling
\begin{equation}
c=c(\Sigma _{c})\left( \frac{M_{c}}{M_{b}}\right) =\allowbreak \left\{
\begin{array}{r}
12.8\text{(}\Sigma _{b}\text{)} \\
10.9\text{(}\Xi _{b}^{\prime }\text{)}%
\end{array}%
\right\} .  \label{cMB}
\end{equation}%
Given $c$ in Eq.(\ref{cMB}), Eqs.(\ref{M1S1}-\ref{M1S2}) lead to the masses
for the S-wave bottom baryons, as listed in Tables I and II,
respectively.

(B) The P-wave baryons($L=1$). A classification that is similar to the
charmed baryons in section III(B) can be given to the baryon state(namely, $%
Q\rightarrow b$). Using the scaling (\ref{scal2}) for the bottom hadrons
where $Q=b$, one can calculate $a_{1}$,
\begin{equation*}
\begin{array}{rr}
\Sigma _{b}: & \ a_{1}(1P)=a_{1}(\Sigma _{c}(1P))=39.4\text{MeV,} \\
\Xi _{b}^{^{\prime }}: & a_{1}(1P)=a_{1}(\Xi _{c}^{^{\prime }}(1P))=33.6%
\text{MeV.}%
\end{array}%
\end{equation*}

For the parameter $a_{2}$ for the singly bottom baryons, the scaling (\ref%
{scal2}) gives, using $a_{2}$ for the SH mesons given in Eq. (\ref{a2SX}),
\begin{equation}
\begin{array}{r}
a_{2}(\Sigma _{b})=a_{2}(\Sigma _{c})\left( \frac{M_{c}}{M_{b}}\right)
=24\left( \frac{1.44}{4.48}\right) =7.7\text{MeV,} \\
a_{2}(\Xi _{b}^{\prime })=a_{2}(\Xi _{c}^{\prime })\left( \frac{M_{c}}{M_{b}}%
\right) =16\left( \frac{1.44}{4.48}\right) =5.1\text{MeV.}%
\end{array}
\label{a2SB}
\end{equation}%
Note that they are smaller than $a_{1}$ roughly by factor of $1/5$. For the $%
\Lambda _{b}/\Xi _{b}$, Eq.(\ref{scal2}) leads to, using $a_{2}$ in the charm sector given in Eq. (\ref{a2charm}),
\begin{equation}
\begin{array}{r}
a_{2}(\Lambda _{b})=a_{2}(\Lambda _{c})\left( \frac{1.44}{4.48}\right)
=\allowbreak 8.61\text{MeV,} \\
a_{2}(\Xi _{b})=a_{2}(\Xi _{c})\left( \frac{1.44}{4.48}\right) =\allowbreak
8.13\text{MeV.}%
\end{array}
\label{a2LB}
\end{equation}

Including the spin-dependent mass splitting, the relation similar to Eq.(\ref%
{M1P2}) yields
\begin{equation}
M(\Lambda _{b}/\Xi _{b},1P)=\bar{M}_{L=1}(\Lambda _{b}/\Xi
_{b})+a_{2}(\Lambda _{b}/\Xi _{b})\left[
\begin{array}{rr}
-1 & 0 \\
0 & 1/2%
\end{array}%
\right] ,  \label{MLX}
\end{equation}%
which gives, combining with the spin-averaged masses in Table V and Eq. (\ref%
{a2LB}), the mass predictions for the P-wave baryons $\Lambda _{b}$ and $\Xi
_{b}$. The results as collected in Table I.

Knowing the coupling strengths $a_{1,2}$ and adding the spin-independent
masses in Table V to the mass splitting (\ref{hsd1})-(\ref{hsd5}), one can
find the masses of the $P$-wave spin multiplets,%
\begin{eqnarray}
M(1/2,0) &=&6009, \\
M(1/2,1) &=&6046-b, \\
M(3/2,1) &=&6050+\frac{1}{2}b, \\
M(3/2,2) &=&6124+\frac{3}{10}b, \\
M(5/2,2) &=&6130-\frac{1}{5}b.
\end{eqnarray}%
for the bottom baryon $\Sigma _{b}^{+}$, and%
\begin{eqnarray}
M(1/2,0) &=&6181  \label{X1} \\
M(1/2,1) &=&6211-b,  \label{X2} \\
M(3/2,1) &=&6216+\frac{1}{2}b,  \label{X3} \\
M(3/2,2) &=&6277+\frac{3}{10}b,  \label{X4} \\
M(5/2,2) &=&6285-\frac{1}{5}b.  \label{X5}
\end{eqnarray}%
for the baryon $\Xi _{b}^{^{\prime }}$, which are plotted against $b$ in
FIG. 3 and FIG. 4, respectively. Note that two adjacent lines in the plots are very close and not distinguishable for the two choices in Eq. (\ref{a2SB}) and Eq. (\ref{a2LB}) and hence only the later case is shown in FIG. 3 and FIG. 4.

In FIG. 3, the observed  mass of the $\Sigma_{b}(6097)^{-}$ pass through the gap between two mass functions $M(3/2,j=1,2)$ while in FIG. 4
the line of the mass function $M(3/2,j=1)$ approachs the mass of
the $\Xi_{b}^{\prime}(6227)^{-}$ at about $b=20$ assuming $b$ is positive. In former case, agreement with data can be achieved most likely when the states $|3/2,j=1,2\rangle$ mixes. In both cases, the spin-parity $J^{P}=3/2^{-}$ is favored for the $\Sigma_{b}(6097)^{-}$ and $\Xi_{b}^{\prime}(6227)^{-}$. The states with $J^{P}=3/2^{-}$ are easier to observe since they decay strongly only via D-wave and are relatively narrow.

(C) The D-wave baryons $\Lambda _{b}/\Xi _{b}$. We compute, as an example,
the masses of the $\Lambda _{b}$(1D). Using the
trajectory parameters in Table V, and $%
M_{b}v_{b}^{2}=4.48(1-4.18^{2}/4.48^{2})=\allowbreak 0.5799$GeV, Eqs. (2-3)
yield a spin-averaged mass for the $D$-wave,
\begin{eqnarray*}
\bar{M}(1D) &=&4.48+\sqrt{2\pi \cdot 0.246+\left[ 0.534+0.5799\right] ^{2}}
\\
&=&6.14927\text{GeV,}
\end{eqnarray*}%
which agrees remarkably well with the observed spin-averaged mass $6149.97$%
MeV of the new resonances $\Lambda _{b}$($3/2^{+},5/2^{+}$) reported very recently by LHCb\cite{Lamdab19}:
\begin{eqnarray*}
M[\Lambda _{b}(6146)^{0}] &=&6146.17\pm 0.33\pm 0.22\pm 0.16\text{MeV,} \\
M[\Lambda _{b}(6152)^{0}] &=&6152.51\pm 0.26\pm 0.22\pm 0.16\text{MeV.}
\end{eqnarray*}
Given the difference $%
5/2$ in $\mathbf{L}\cdot \mathbf{S}_{Q}$ between the $5/2^{+}$ and $3/2^{+}$
states, one can extract $a_{2}$ for the $\Lambda _{c}(1D)$ to have $%
a_{2}[\Lambda _{c}(1D)]=\frac{2}{5}[\Lambda _{c}(5/2^{+})-\Lambda
_{c}(3/2^{+})]=10.21$MeV. The scaling relation (\ref{scal2}) gives then
\begin{equation*}
a_{2}[\Lambda _{b}(1D)]=a_{2}[\Lambda _{c}(1D)]\left( \frac{1.44}{4.48}%
\right) =3.28\text{MeV,}
\end{equation*}%
which predicts(in MeV), by Eq. (\ref{HDw}),
\begin{eqnarray*}
M[\Lambda _{b}(1D)] &=&6149.3+3.28\left[
\begin{array}{rr}
-3/2 & 0 \\
0~~~ & 1%
\end{array}%
\right]  \\
&=&\left[
\begin{array}{rr}
6144.4 & 0~~~~ \\
0~~~~ & 6152.6%
\end{array}%
\right] .
\end{eqnarray*}%
corresponding to $J= {3/2^{+},5/2^{+}}$, respectively. This is in good agreement  with the observations.

Due to the lacking of the $B/B_{s}$ meson data for D-wave unfortunately, we
are not able to test the scaling relation (\ref{res}) directly with the SH mesons. An alternative is to verify it via scaling from the $D(1D)$ mesons to the $\Lambda _{c}(1D)$ and then to the $\Lambda _{b}(1D)$. Combined with Eq.(\ref{scaling}) and (C23) in Appendix C, Eq. (\ref{res}) yields

\begin{eqnarray}
a_{2}[\Lambda _{b}] &=&a_{2}[\Lambda _{c}(1D)]\left( \frac{M_{c}}{M_{b}}%
\right) =\frac{a_{2}[D(1D)]}{1+m_{d}/M_{c}}\left( \frac{M_{c}}{M_{b}}\right)
\notag \\
&=&\frac{12.6\text{MeV}}{1+0.535/1.44}\left( \frac{1.44}{4.48}\right)
=\allowbreak 2.95\text{MeV},
\label{aLB}
\end{eqnarray}%
which gives%
\begin{eqnarray*}
M[\Lambda_{b}(1D)] &=&6149.27+\{-3/2,1\}\ast \allowbreak 2.95 \\
&=&\left\{ 6144.8,6152.2\right\} \text{MeV.}
\end{eqnarray*}%
This agrees with the LHCb observations equally well. The same procedure as in Eq. (\ref{aLB}) applies for computing the D-wave masses of the $\Xi_{b}$, with the results collected in Table I.

Finally, we explain why the some of the four observed SH
baryons discussed in this work can not be the lower $J^{P}=1/2^{-}$
multiplet in the P-wave SH baryons so that they can be seen easily
experimentally comparing with the $J^{P}=3/2^{-}$ multiplets. In the
following, we choose $\Xi _{b}^{\prime }(6227)^{-}$ as a example to address
this question briefly in the light of strong decay behavior which was not the main topics of this work. For more discussions on decays of the $%
\Sigma _{b}(6097)/\Xi _{b}^{^{\prime }}(6227)^{-}$, see the literatures \cite{QMao,PYang:D2019,WangLZ:D19,Cui:2019dzj}.

First of all, we note that experiments by LHCb using $pp$ collision data found the $\Xi _{b}^{\prime }(6227)^{-}$ to be relatively narrow\cite{Aaij et al:D19,Aaij:D18}
\begin{eqnarray*}
\Gamma \lbrack \Xi _{b}^{\prime }(6227)^{-}] &=&18.1\pm 5.4\pm 1.8\text{MeV,}
\end{eqnarray*}%

In principle the $1/2^{-}$ assignment for the $\Xi _{b}^{\prime }(6227)^{-}$
is allowed since in the P-wave the diquark angular momentum can be $%
j=1\oplus 1=\{0,1,2\}$ and couple with $S_{Q}=1/2$ to give $J=1/2$. If this
is the case, the decay to the $1/2^{+}$ ground-state $\Xi_{b}\pi ^{-}$(or $%
\Lambda_{b}K^{-}$) is very likely to happen via the $S$-wave by absorbing
the $L=1$ orbital motion into the diquark system to flip the spin of one
quark in diquark $bs^{\uparrow }q^{\uparrow }(1P)\rightarrow bs^{\uparrow
}q^{\downarrow }(1S_{1/2})+q^{\uparrow }\bar{q}^{\downarrow }$(S-wave). This
yields normally a large decay width, contradictory with the relatively
narrow width of the $\Xi_{b}^{\prime }(6227)$ shown above. For the $%
J^{P}=3/2^{-}$ SH baryons in P-wave, this kind of absorbing is prohibited by
the angular momentum conservation, and the P-wave decay is also banned by
the parity. So the $\Xi_{b}^{\prime }(6227)$ can decay strongly only via $D$-wave
which agrees with its relatively narrow width due to the higher $L$
suppression. So, unlike the $1/2^{-}$ state, a $3/2^{-}$ baryon $\Xi _{b}$
in P-wave is easier to be seen in experiments due to its relatively narrow
width. In addition, though the $5/2^{-}$ assignment for the $\Sigma _{b}(6097)$ seems to be marginal, it can not be excluded fully in the light of our mass analysis.

\begin{figure}[h]
\centering
\begin{tabular}{c}
\includegraphics[width=0.44\textwidth]{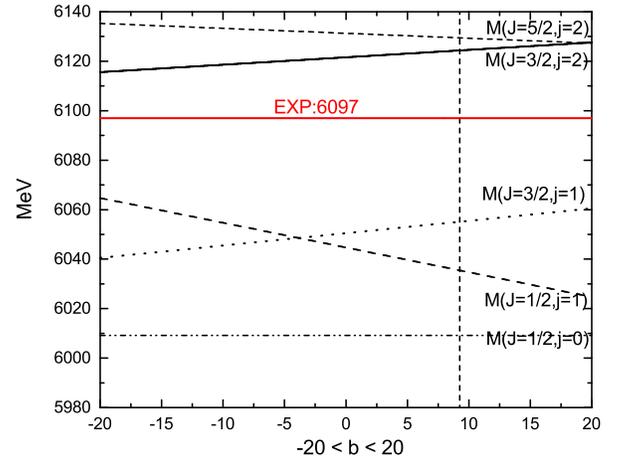}
\end{tabular}
\caption{The masses of the $P$-wave $\Sigma _{b}$ against the parameter $b$.}\label{potentialshape}
\end{figure}
\begin{figure}[h]
\centering
\begin{tabular}{c}
\includegraphics[width=0.44\textwidth]{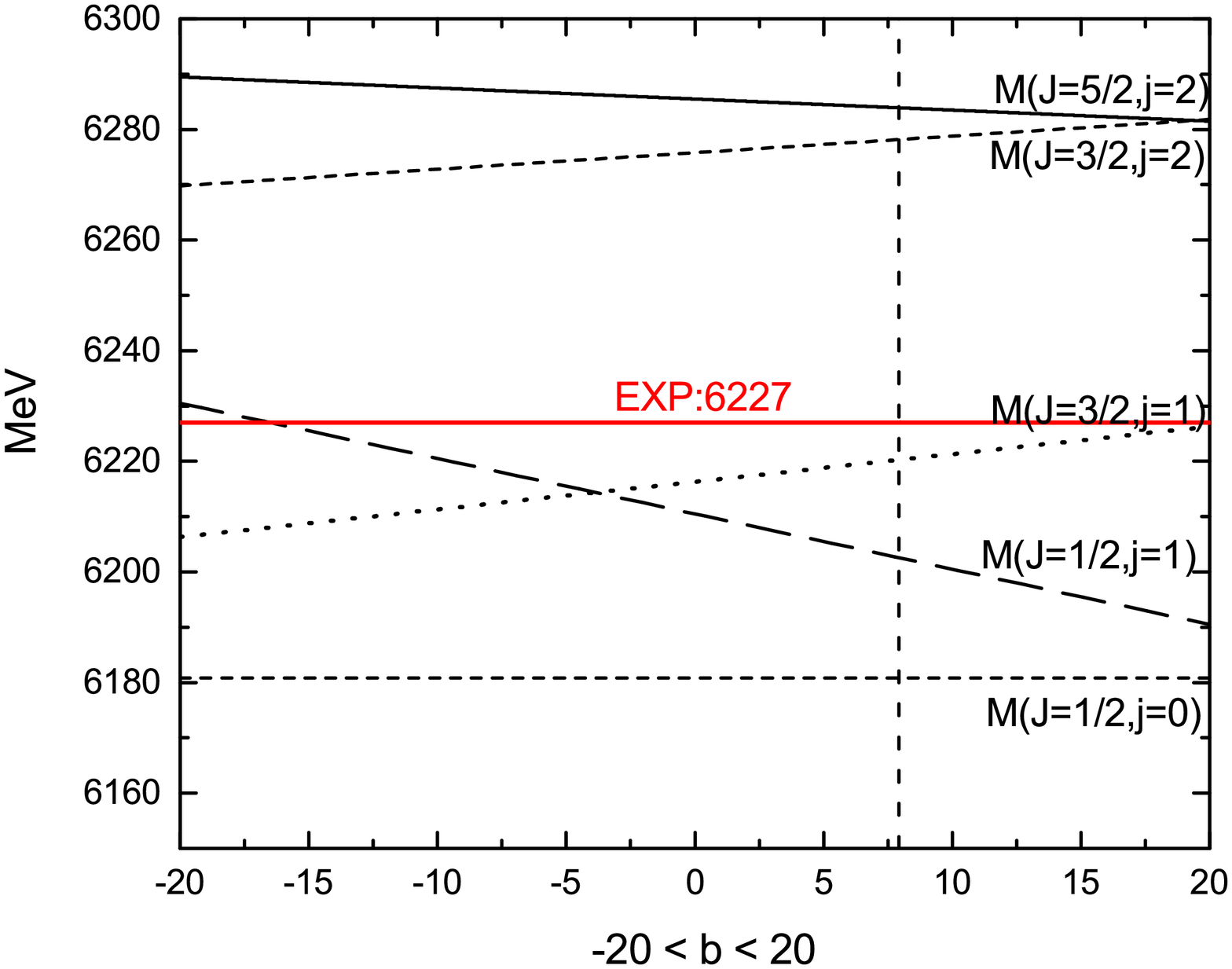}
\end{tabular}
\caption{The masses of the $P$-wave $\Xi_{b}^{\prime}$ against the parameter $b$.}\label{potentialshape}
\end{figure}

\section{Summary and discussions}\label{sec5}
In this work we use Regge approach in the heavy quark-diquark picture
to re-examine the orbitally excited spectrum of the charmed and bottom
baryons. Our analysis via a simple linear Regge relation (\ref{GRegg}) with
the baryon mass shifted by the heavy quark mass $M\rightarrow M-M_{Q}$
indicates that the relation is able to describe the spin-averaged mass
spectrum of the orbitally-excited charmed and bottom baryons($\Lambda _{Q}$,
$\Xi _{Q}$, $\Sigma _{Q}$ and $\Xi _{Q}^{\prime }$, $Q=c,b$). The further
computations on spin-dependent mass splitting suggest that the baryons $\Xi
_{b}(6227)^{-}$ and the $\Sigma _{b}(6097)^{-}$, and the $\Sigma _{c}(2800)$
and the $\Xi _{c}^{\prime }(2930)$ are all the $1P$-wave baryons with the
spin-parity $J^{P}=3/2^{-}$ preferably. We also present mass predictions of
the unseen bottom baryon $\Xi _{b}$ in their $1P$ and $1D$ states, providing
some useful clues for the coming experiments to find them, especially the
LHCb.

It is noticeable that the relation (\ref{GRegg}) is derived from the
classical relativistic energy and angular momentum of the rotating QCD
string model, with no attempt made to obtain quantum correction to it.
However, this relation shows the main feature predicted by the quantized
theory of the QCD string \cite{Rebbi:PR74,LaCourseO:D89,SW:jh18,BakerSD:02}:
the linear functional form of trajectory remains intact, only with the
intercept and the hadron mass shifted. When
taking $L$ to be the quantized numbers, the approach in this work can be
viewed, in a sense, as an leading semi-classical approximation of quantized
theory of string. A recent quantum treatment of rotating string\cite{SW:jh18}
in the nonvanishing endpoint-mass case yields a similar conclusion. We also note that for a string tied to one heavy and one light quarks, the mass shift $M\rightarrow M-M_{Q}$ in the relation (\ref{GRegg}) is crucial for describing the moderate nonlinear behavior of the baryon mass squared $M^{2}$ as a function of $L$\cite{LaCourseO:D89}(see FIGs.6-7) at low-$L$: such a nonlinearity is absorbed into Eq. (\ref{GRegg}) largely, since the later takes the form:
\begin{equation}
M^{2}-2M_{Q}M+M_{Q}^{2}=\pi aL+a_{0},  \label{EE0}
\end{equation}%
when expanding it in $M_{Q}$. Obviously, a nonlinear dependence in $M^{2}$ happens in the LHS of Eq. (\ref{EE0}) which tends to vanish when L is large. This is in consistent with the numerical computations in Ref.\cite{LaCourseO:D89},
and is supported by a systematic examination\cite{KChenL:C18} for
whole SH hadron systems (the SH mesons and baryons).

In order to compute the spin-dependent mass, we employed the scaling relations
based on similarity in the color-configurations between a SH baryon and its
meson partner or among the heavy flavor partners. While they are of approximate
these relations turn out to be valuable as they have exhibited the most
general features of the spin-dependent relativistic corrections, which scale as the magnetic moment $\mathbf{S}_{d}/m_{d}$ for the $a_{1}$-term and as $\mathbf{S}_{Q}/M_{Q}$ for the $a_{2}$-term in Eq.(\ref{Hsd}). We note that the color-structure similarity among these SH hadrons is implicitly assumed in the quark-diquark picture.

By the way, the following remarks are in order:

(1) Without adding or tunning of the parameters, a simple application of our computational procedure to the D-wave bottom baryon $\Lambda_{b}$ yields a spin-averaged mass and spin-dependent mass splitting which agree well with the observed spin-averaged mass $6149.97$MeV of the newly reported resonances $\Lambda _{b}$ \cite{Lamdab19} and its mass splitting $-3.80$MeV and $2.54$MeV for the $3/2^{+}$ and $5/2^{+}$ states, respectively.

(2) By predicting spin-averaged masses, the relation (\ref{GRegg}) are helpful in itself in identifying the state order of the excited spin-multiplets in the sense of the bisection of the mass intervals. To accommodate them further, detailed computations about the mass differences of the spin-multiplets are necessary.

(3) Similar decay behaviors (the pseudoscalar meson emission and relatively narrow widths) seen for the baryons $\Xi
_{b}(6227)^{-}$/$\Sigma _{b}(6097)^{-}$, the $\Sigma _{c}(2800)$
/$\Xi_{c}^{\prime }(2930)$ prefer to support that they are all the $1P$-wave baryons with $J^{P}=3/2^{-}$. Further exploring of the parameter $b$ for them becomes valuable for decoding inner-structures of the P-wave partners of these baryons.

(4) The method in this work can in principle be used to estimate the higher
orbital excitations of the bottom baryons, with the experimental mass information of the excited meson partners or the charmed partners increasing. To go beyond the limitation having to refer to the SH meson mass spectrum in computing the mass splitting, the quantized string dynamics is desirable for further exploring the heavy baryon spectroscopy.

\emph{Notes added.--}After submission of this work, two newly reported resonances $\Lambda _{b}(6146)^{0}$ and the $\Lambda _{b}(6146)^{0}$\cite{Lamdab19} come to our attention, which can be nicely accommodated as the D-wave singly bottom baryons using our approach in this work. The agreement of our prediction with the observed mass data is remarkable.

\section*{ACKNOWLEDGMENTS}

D. J. thanks H-X. Chen for useful discussions. D. J. is supported by the National Natural Science Foundation of China under the no. 11565023 and the Feitian Distinguished Professor Program of Gansu(2014-2016). D. J. and A. H. are supported in part by Grant-in-Aid for Scientific Research on Innovative Areas, \textquotedblleft Clustering as a window on the hierarchical structure of quantum
systems \textquotedblright.

\section*{APPENDIX A}\vspace{0.5cm}

For the orbitally excitations, the classical energy and orbital angular
momentum for the rotating QCD string read \cite%
{JohnsonN:PRD79,SelemW06,LaCourseO:D89}
\begin{equation}
E=\frac{m_{Q}}{\sqrt{1-v_{Q}^{2}}}+\frac{m}{\sqrt{1-v_{d}^{2}}}+\frac{a}{%
\omega }\sum_{i=Q,d}\int_{0}^{v_{i}}\frac{du}{\sqrt{1-u^{2}}},  \tag{A1}
\end{equation}%
\begin{equation}
L=\frac{m_{Q}v_{Q}^{2}/\omega }{\sqrt{1-v_{Q}^{2}}}+\frac{mv_{d}^{2}/\omega
}{\sqrt{1-v_{d}^{2}}}+\frac{a}{\omega ^{2}}\sum_{i=Q,d}\int_{0}^{v_{i}}\frac{%
u^{2}du}{\sqrt{1-u^{2}}},  \tag{A2}
\end{equation}%
where $m_{Q}$ and $m$ are the bare masses of the heavy quark $Q$ and light
diquarks $d$ and $v_{i}=\omega r_{i}$($i=Q,d$) the velocity of the string
end tied to the quark $i=Q,d$. It is known that the Selem-Wilczek relation (%
\ref{SW}) follows from upon the ($m_{i}\omega /a$)-expansion of (A1) and (A2) \cite{SelemW06}. Following \cite{BChen:A15}, we define the
effective (dynamical) masses of the heavy quark and the light diquark in the
CM frame of the baryon by
\begin{equation}
M_{Q}=\frac{m_{Q}}{\sqrt{1-v_{Q}^{2}}},m_{d}=\frac{m}{\sqrt{1-v_{d}^{2}}},\tag{A3}
\end{equation}%
to rewrite (A1) and (A2) as
\begin{equation}
E=M_{Q}+m_{d}+\frac{a}{\omega }[\arcsin (v_{d})+\arcsin (v_{Q})],  \tag{A4}
\end{equation}
\begin{equation}
L=\frac{1}{\omega }(M_{Q}v_{Q}^{2}+m_{d}v_{d}^{2})+\frac{a}{2\omega ^{2}}%
\sum_{i=Q,d}\left[ \arcsin (v_{i})-v_{i}\sqrt{1-v_{i}^{2}}\right] ,  \tag{A5}
\end{equation}%
in which the last term in (A4) is the contribution to the orbital angular
momentum due to the string rotating.

The boundary condition of string at ends with heavy quark gives
\begin{equation}
\frac{a}{\omega }=\frac{m_{Q}v_{Q}}{1-v_{Q}^{2}}=\frac{M_{Q}v_{Q}}{\sqrt{%
1-v_{Q}^{2}}},  \tag{A6}
\end{equation}%
or
\begin{equation}
\frac{a}{\omega }\simeq P_{Q}+\frac{P_{Q}^{3}}{2M_{Q}^{2}}\text{,}  \tag{A7}
\end{equation}%
with $P_{Q}\equiv M_{Q}v_{Q}$ the conserved momentum of the heavy quark.
Expanding Eqs. (A3) and (A4) in $v_{Q}$ and $m$ gives, up to $v_{Q}^{4}$ and
the leading order of $m$,
\begin{equation}
E =M_{Q}+m_{d}+\frac{\pi a }{2\omega }+\frac{a}{\omega }\left[ v_{Q}-%
\frac{m}{m_{d}}+\frac{1}{6}v_{Q}^{3}\right] +\mathcal{O}[v_{Q}^{5}],
\tag{A8}
\end{equation}
\begin{equation}
\omega L =m_{d}+M_{Q}v_{Q}^{2}+\frac{a}{\omega }\left[ \frac{\pi }{4}-%
\frac{m}{m_{d}}\right] +\frac{a}{3\omega }v_{Q}^{3}+\mathcal{O}%
[v_{Q}^{5}].  \tag{A9}
\end{equation}

 Using Eq. (A7) and upon eliminating $\omega $, Eqs.(A8)-(A9) combines to give, when ignoring the small term $m/m_{d}$,
\begin{equation}
(E-M_{Q})^{2}=\pi aL+\left( m_{d}+\frac{P_{Q}^{2}}{M_{Q}}\right)
^{2}-2mP_{Q}.  \tag{A10}
\end{equation}
This leads to Eq. (\ref{GRegg}) when taking the bare mass $m\rightarrow 0$. By the way, we note that Eq. (\ref{PQ}) follows by rewriting the velocity $
v_{Q}^{2}$=$1-(m_{Q}/M_{Q})^{2}$, with $m_{Q}\equiv$$m_{bareQ}$ the bare mass of the heavy quark $Q$.

Note that the heavy quark limit(HQL) is defined as the largeness of the bare mass $m_{Q}$ of the heavy quark $Q$ entering the QCD Lagrangian while its 3-momentum $P_{Q}=M_{Q}v_{Q}$ remains fixed as Q behaves like a static external source\cite{Manohar:D07} within the bound state of hadrons. It follows from Eq. (A3) that in the HQL the effective mass $M_{Q}$ tends to infinity also while the 3-velocity $v_{Q}$ tends to vanish, that is, the heavy quarks move nonrelativistically within hadrons, as it should be. One can readily check that in the HQL, Eqs. (A8) and (A9) become $E =M_{Q}+m_{d}+\frac{\pi a}{2\omega}$ and $\omega L =m_{d}+\frac{a \pi}{4\omega}$, which yield $(E-M_{Q})^{2}=\pi aL+m_{d}^{2}$. Here, $M_{Q}v_{Q}^{2}=P_{Q}^{2}/M_{Q}$ tends to vanish in the RHS of Eq. (A9).

\section*{APPENDIX B}

For the heavy quark-diquark systems, the matrix elements of  $\mathbf{L}\cdot
\mathbf{S}_{i}$ can be calculated by explicit construction of the baryon
states with a given $J_{3}$ as linear combinations of the states $%
|S_{d3},S_{Q3},L_{3}\rangle $ with $S_{d3}+S_{Q3}+L_{3}=J_{3}$. Due to the rotation invariance of the matrix elements, it suffices to use a single $%
J_{3}$ for each and, one can use
\begin{equation*}
\mathbf{L}\cdot \mathbf{S}_{i}=\frac{1}{2}\left[ L_{+}S_{i-}+L_{-}S_{i+}%
\right] +L_{3}S_{i3}, \tag{B1}
\end{equation*}%
to find their elements by applying $\mathbf{L}\cdot \mathbf{S}_{i}$ on the
third components of angular momenta. In the rep. of $|L_{z},S_{Qz},S_{dz}%
\rangle $ these component are given by the following basis states \cite{KarlinerR:D18}
\begin{equation*}
\begin{split}
|1^{2}P_{1/2},J_{z}=1/2\rangle =&\frac{\sqrt{2}}{3}|1,-1/2,0\rangle -\frac{1}{3%
}|0,1/2,0\rangle \\ &-\frac{\sqrt{2}}{3}|0,-1/2,1\rangle +\frac{2}{3}%
|-1,1/2,1\rangle,
\end{split}
\end{equation*}
\begin{eqnarray*}
\begin{split}
|1^{4}P_{1/2},J_{z} =1/2\rangle =&\frac{1}{\sqrt{2}}|1,1/2,-1\rangle -\frac{%
1}{3}|1,-1/2,0\rangle \\ &-\frac{\sqrt{2}}{3}|0,1/2,0\rangle +\frac{1}{3}
|0,-1/2,1\rangle \\ &+\frac{1}{3\sqrt{2}}|-1,1/2,1\rangle,
\end{split}
\end{eqnarray*}%
\begin{equation*}
|1^{2}P_{3/2},J_{z}=3/2\rangle =\sqrt{\frac{2}{3}}|1,-1/2,1\rangle -\sqrt{%
\frac{1}{3}}|0,1/2,1\rangle,
\end{equation*}%
\begin{equation*}
\begin{split}
|1^{4}P_{3/2},J_{z}=3/2\rangle =& \sqrt{\frac{3}{5}}|1,1/2,0\rangle -
\sqrt{\frac{2}{15}}|1,-1/2,1\rangle \\ &-\frac{2}{\sqrt{15}}|0,1/2,1\rangle,
\end{split}
\end{equation*}   
\begin{equation*}
|1^{4}P_{5/2},J_{z}=5/2\rangle =|1,1/2,1\rangle.
\end{equation*}%

Having these component states (in $L-S$ coupling), one can compute the matrix elements of $\mathbf{L}\cdot \mathbf{S}_{d}$,
\begin{equation*}
\left\langle \mathbf{L}\cdot \mathbf{S}_{d}\right\rangle _{J=\frac{1}{2}}=%
\left[
\begin{array}{rr}
-\frac{4}{3} & -\frac{\sqrt{2}}{3} \\
-\frac{\sqrt{2}}{3} & -\frac{5}{3}%
\end{array} %
\right],    \tag{B1}
\end{equation*}%
\begin{equation*}
\left\langle \mathbf{L}\cdot \mathbf{S}_{d}\right\rangle _{J=\frac{3}{2}}=%
\left[
\begin{array}{rr}
\frac{2}{3} & -\frac{\sqrt{5}}{3} \\
-\frac{\sqrt{5}}{3} & -\frac{2}{3}%
\end{array} %
\right].    \tag{B2}
\end{equation*}%

Accordingly, the states in terms of the $J-j$ coupling can be expressed as linear
combinations of states with definite $L-S$ coupling (for each eigenvalue $\lambda$ of $\mathbf{L}\cdot \mathbf{S}_{d}$.) \cite{KarlinerR:D18}:
\begin{equation}
\lambda=-2:|J=\frac{1}{2}, j=0\rangle=\sqrt{\frac{1}{3}}|1^{2}P_{1/2}\rangle+\sqrt{\frac{2}{3}}|1^{4}P_{1/2}\rangle\tag{B3},
\end{equation}
\begin{equation}
\lambda=-1:|J=\frac{1}{2}, j=1\rangle=\sqrt{\frac{2}{3}}|1^{2}P_{1/2}\rangle-\sqrt{\frac{1}{3}}|1^{4}P_{1/2}\rangle\tag{B4},
\end{equation}
\begin{equation}
\lambda=-1:|J=\frac{3}{2}, j=1\rangle=\sqrt{\frac{1}{6}}|1^{2}P_{3/2}\rangle+\sqrt{\frac{5}{6}}|1^{4}P_{3/2}\rangle\tag{B5},
\end{equation}
\begin{equation}
\lambda=+1:|J=\frac{3}{2}, j=2\rangle=\sqrt{\frac{5}{6}}|1^{2}P_{3/2}\rangle-\sqrt{\frac{1}{6}}|1^{4}P_{3/2}\rangle\tag{B6},
\end{equation}
\begin{equation}
\lambda=+1:|J=\frac{5}{2}, j=2\rangle=|1^{4}P_{5/2}\rangle\tag{B7},
\end{equation}
which can be used to average $\mathbf{L}\cdot\mathbf{S}_{Q}$ and $\mathbf{S}_{12}$. The detailed results are collected in Table X.

\section*{APPENDIX C}

We describe here some of the details on how to estimate the values of the parameters {$a_{1},a_{2}, b, c$} that enter the spin-dependent interaction (\ref{Hsd}) for the SH mesons where $\mathbf{S}_{d}$ is replaced everywhere in Appendix C by the spin $\mathbf{S}_{q}$ of light quark. The following three cases of excited states are considered in order:
\medskip

(1) The S-wave SH mesons. In this ground-state case, only the contact term $c\mathbf{S}_{Q}\cdot \mathbf{S}_{q}$
remains in the spin-dependent interaction (\ref{Hsd}). Since the expectation
values of $\mathbf{S}_{Q}\cdot \mathbf{S}_{q}$ are $-3/4$ and $1/4$ in the $%
1^{1}S_{0}$ and $1^{3}S_{1}$ meson states, respectively, the parameter $c$
can be estimated simply by confronting the predicted mass difference between two S-states, $c(1/4+3/4)=c$, with the measured S-wave masses of the $D$ mesons in the Table VI. The result is%
\begin{equation}
c=(2010.3-1869.7)\text{MeV}=140.6\text{MeV.}  \tag{C1}
\end{equation}
\medskip

(2) The P-wave SH mesons. Similar to Appendix B, a calculation can be
performed for the P-wave SH mesons. As in the case of the SH baryons, we find it convenient to work in the $j-j$ basis in which
the analogue of the first term in Eq.(\ref{Hsd}), $a_{1}\mathbf{L}%
\cdot \mathbf{S}_{q}$, is diagonal. We first calculate the expectation
values of $\mathbf{L}\cdot \mathbf{S}$ and the tensor operator $\mathbf{S}%
_{12}$ in the basis states $^{2S+1}P_{J}$. The results are shown in Table XI.

\medskip
\renewcommand\tabcolsep{0.68cm}
\renewcommand{\arraystretch}{1.5}
\begin{table}[!htbp]
\caption{Expectation values of the spin-dependent terms for P-wave $c\bar{q%
}$ mesons in the $L-S$-coupling basis states.}
\begin{tabular}{cccc}
\hline\hline
State & $\left\langle \mathbf{L}\cdot \mathbf{S}\right\rangle $ & $%
\left\langle \mathbf{S}_{12}\right\rangle $ & $\left\langle \mathbf{S}%
_{Q}\cdot \mathbf{S}_{q}\right\rangle $ \\ \hline
$^{3}P_{0}$ & $-2$ & $-4$ & $1/4$ \\
$^{1}P_{1}$ & $0$ & $0$ & $-3/4$ \\ \hline
$^{3}P_{1}$ & $-1$ & $2$ & $1/4$ \\
$^{3}P_{2}$ & $1$ & $-2/5$ & $1/4$ \\ \hline\hline
\end{tabular}
\end{table}

To evaluate the expectation values of $\mathbf{L}\cdot \mathbf{S}_{i}$($i=q,Q
$), it is convenient to use the $L-S$ basis. When labelling the states by $%
|S_{q3},S_{Q3},L_{3}\rangle $, one can construct the $L-S$ basis by

\begin{equation}
|^{3}P_{0},J_{3} =0\rangle =\left\vert -\frac{1}{2},-\frac{1}{2}%
,1\right\rangle , \tag{C2}
\end{equation}
\begin{equation}
|^{1}P_{1},J_{3} =1\rangle =\frac{1}{\sqrt{2}}\left\vert \frac{1}{2},-%
\frac{1}{2},1\right\rangle -\frac{1}{\sqrt{2}}\left\vert -\frac{1}{2},\frac{1%
}{2},1\right\rangle ,  \tag{C3}
\end{equation}
\begin{equation}
|^{3}P_{1},J_{3} =1\rangle =\frac{1}{2}\left\vert -\frac{1}{2},\frac{1}{2}%
,1\right\rangle +\frac{1}{2}\left\vert \frac{1}{2},-\frac{1}{2}%
,1\right\rangle -\frac{1}{\sqrt{2}}\left\vert \frac{1}{2},\frac{1}{2}%
,0\right\rangle ,\\  \tag{C4}
\end{equation}
\begin{equation}
|^{3}P_{2},J_{3} =2\rangle =\left\vert \frac{1}{2},\frac{1}{2}%
,1\right\rangle .  \tag{C5}
\end{equation}%
Using (B1) in the $J=1$ subspace given by the basis ($^{1}P_{1},^{3}P_{1}$), the
matrices describing mixing of the $J=1$ states are
\begin{equation*}
\left\langle \mathbf{L}\cdot \mathbf{S}_{q}\right\rangle _{J=1}=%
\left[
\begin{array}{rr}
0~~~& 1/\sqrt{2} \\\tag{C6}
1/\sqrt{2} & -1/2%
\end{array} %
\right],
\end{equation*}%
\begin{equation*}
\left\langle \mathbf{L}\cdot \mathbf{S}_{Q}\right\rangle _{J=1}=%
\left[
\begin{array}{rr}
0~~~& -1/\sqrt{2} \\\tag{C7}
-1/\sqrt{2} & -1/2%
\end{array} %
\right].
\end{equation*}%
The first matrix, which has the eigenvalues $-1$ and $1/2$, can be
diagonalized by transforming the $L-S$ eigenstates to the $|J,j\rangle $
eigenstates. This transformation, when written in the whole space, can be given by%
\begin{eqnarray*}
|J &=&0,j=1/2\rangle =|^{3}P_{0}\rangle , \\
|J &=&1,j=1/2\rangle =\sqrt{\frac{1}{3}}|^{1}P_{1}\rangle -\sqrt{\frac{2}{3}}%
|^{3}P_{1}\rangle , \\
|J &=&1,j=3/2\rangle =\sqrt{\frac{2}{3}}|^{1}P_{1}\rangle +\sqrt{\frac{1}{3}}%
|^{3}P_{1}\rangle , \\
|J &=&2,j=3/2\rangle =|^{3}P_{2}\rangle .
\end{eqnarray*}

With the expressions above one can work out the expectation values of all
spin-dependent operators contributing to the masses of the P-wave $Q\bar{q}$
mesons and thereby obtain estimations for the spin-dependent mass splitting
via expanding around the eigenstates of $\mathbf{j}=\mathbf{L}+\mathbf{S}_{q}
$. Including the spin-averaged mass $\bar{M}$, the results for the P-wave are
\cite{KarlinerR:D15},
\begin{equation*}
M(0,1/2) = \bar{M}-a_{1}-a_{2}-b-\frac{c}{4},  \\ \tag{C8}
\end{equation*}
\begin{equation*}
M(1,1/2) = \bar{M}-a_{1}+\frac{1}{3}a_{2}+\frac{1}{3}b-\frac{c}{12}, \tag{C9}
\end{equation*}
\begin{equation*}
M(1,3/2) = \bar{M}+\frac{1}{2}a_{1}-\frac{5}{6}a_{2}+\frac{1}{6}b-\frac{5c}{%
12},  \\ \tag{C10}
\end{equation*}
\begin{equation*}
M(2,3/2) = \bar{M}+\frac{1}{2}a_{1}+\frac{1}{2}a_{2}-\frac{1}{10}b+\frac{c}{%
4}.  \\ \tag{C11}
\end{equation*}%
Given Eqs. (C8-C11), the parameters ($a_{1},a_{2},b$) which reproduce the observed masses of the $c\bar{s}$ mesons in Table VI are\cite{KarlinerR:D15}
\begin{equation}
\begin{Bmatrix}
\bar{M}\text{(MeV),} & a_{1}\text{(MeV),} & a_{2}\text{(MeV)} & b\text{(MeV)}
\\
2513.4, & 89.4, & 40.7, & 65.6,%
\end{Bmatrix}
\tag{C12}
\end{equation}%
where the parameter $c$ in the hyperfine term is taken to be zero as it is quite
small in the P-wave state.
\medskip

(3) The D-wave SH mesons. The similar procedure applies for the D-wave $Q%
\bar{q}$ mesons. Using the state $|S_{q3},S_{Q3},L_{3}\rangle $, one can
explicitly construct the $L-S$ basis for the D-wave $Q\bar{q}$ states by

\begin{equation*}
|^{3}D_{1},J_{3} =1\rangle =\sqrt{\frac{3}{5}}\left\vert -\frac{1}{2},-%
\frac{1}{2},2\right\rangle +\sqrt{\frac{1}{10}}\left\vert \frac{1}{2},\frac{1%
}{2},0\right\rangle\\
\end{equation*}
\begin{equation}
-\sqrt{\frac{3}{20}}\left\vert \frac{1}{2},-\frac{1}{2},1\right\rangle -%
\sqrt{\frac{3}{20}}\left\vert -\frac{1}{2},\frac{1}{2},1\right\rangle,
\tag{C13} \\
\end{equation}
\begin{equation}
|^{1}D_{2},J_{3} =  2\rangle =\frac{1}{\sqrt{2}}\left\vert \frac{1}{2},-%
\frac{1}{2},2\right\rangle -\frac{1}{\sqrt{2}}\left\vert -\frac{1}{2},\frac{1%
}{2},2\right\rangle ,\tag{C14} \\
\end{equation}
\begin{equation}
|^{3}D_{2},J_{3} = 2\rangle =-\frac{1}{\sqrt{3}}\left\vert \frac{1}{2},%
\frac{1}{2},1\right\rangle +\frac{1}{\sqrt{3}}\left\vert \frac{1}{2},-\frac{1%
}{2},2\right\rangle +\frac{1}{\sqrt{3}}\left\vert -\frac{1}{2},\frac{1}{2}%
,2\right\rangle , \tag{C15} \\
\end{equation}
\begin{equation}
|^{3}D_{3},J_{3} =3 \rangle =\left\vert \frac{1}{2},\frac{1}{2}%
,2\right\rangle . \tag{C16}
\end{equation}

Utilizing (B1) again in the $J=2$ subspace given by the basis ($%
^{1}D_{2},^{3}D_{2}$), the matrices describing mixing of the $J=2$ states
are

\begin{equation*}
\left\langle \mathbf{L}\cdot \mathbf{S}_{q}\right\rangle _{J=2} =\left[
\begin{array}{rr}
0~~~~ &\sqrt{3/2} \\\tag{C17}
\sqrt{3/2} & -1/2
\end{array}%
\right] ,
\end{equation*}
\begin{equation*}
\left\langle \mathbf{L}\cdot \mathbf{S}_{Q}\right\rangle _{J=2} =\left[
\begin{array}{rr}
0~~~~ & -\sqrt{3/2} \\\tag{C18}
-\sqrt{3/2} & -1/2
\end{array}
\right] .
\end{equation*}
The first matrix, which has the eigenvalues $-3/2$ and $1$, can be
diagonalized by the following basis transformation between the $L-S$ eigenstates
and the $|J,j\rangle $ eigenstates. In the whole space, the transformation takes the form:
\begin{equation*}
|J =1,j=3/2\rangle =|^{3}D_{1}\rangle , \\
\end{equation*}
\begin{equation*}
|J =2,j=3/2\rangle =-\sqrt{\frac{2}{5}}|^{1}P_{2}\rangle -\sqrt{\frac{3}{5}%
}|^{3}P_{2}\rangle , \\
\end{equation*}
\begin{equation*}
|J =2,j=5/2\rangle =\sqrt{\frac{3}{5}}|^{1}P_{2}\rangle +\sqrt{\frac{2}{5}}%
|^{3}P_{2}\rangle , \\
\end{equation*}
\begin{equation*}
|J =3,j=5/2\rangle =|^{3}P_{3}\rangle .
\end{equation*}
Given these relations, the calculations for the expectation values of all
mass splitting operators can be performed and the mass expression for the
D-wave $Q\bar{q}$ mesons can be obtained via the procedure similar to that to have Eqs.(\ref{hsd1})-(\ref{hsd5}). The results are
\begin{equation*}
M(1,3/2) =\bar{M}-\frac{3}{2}a_{1}-\frac{3}{2}a_{2}-\frac{5}{2}b+\frac{c}{%
4},   \\ \tag{C19}
\end{equation*}
\begin{equation*}
M(2,3/2) =\bar{M}-\frac{3}{2}a_{1}+\frac{9}{10}a_{2}+\frac{3}{2}b-\frac{3c%
}{2},   \\ \tag{C20}
\end{equation*}
\begin{equation*}
M(2,5/2) =\bar{M}+a_{1}-\frac{7}{5}a_{2}+\frac{7}{5}b-\frac{7c}{20},
 \\ \tag{C21}
\end{equation*}
\begin{equation*}
M(3,5/2) =\bar{M}+a_{1}+a_{2}-b-\frac{c}{4}.
 \tag{C22}
\end{equation*}

In the D-wave case, only two observed masses are available in Table VI for
each family of the $D/D_{s}$ mesons. Since only the mass splitting $\Delta M=M-\bar{M}$ is relevant here, we use the mass splitting $\Delta M$ given by the quark model(QM) predictions\cite{EFG:C10} rescaled by the deviation factor $2.04$ so that the rescaled mass splitting match the observed data. The details are listed in Table XII.

\renewcommand\tabcolsep{0.2cm}
\renewcommand{\arraystretch}{1.5}
\begin{table}
\caption{The masses and corresponding mass splitting $\Delta M=M-\bar{M}$
for the D-wave $D$ mesons. The QM stands for the quark model predictions($%
\bar{M}=2834.3$MeV). The rescaled mass splitting are obtained by
multiplying $\Delta M$(QM) by a deviation factor $(2763.5-2737)/(2863-2850)=2.04$.}
\begin{tabular}{ccccc}
\hline\hline
States & $M$(Exp.) & $M$(QM)\cite{EFG:C10} & $\Delta M$(QM) & $\Delta M$%
(rescaled) \\ \hline
$|1,3/2\rangle $ & -- & $2788$ & $-46.3$ & $-94.4$ \\
$|2,3/2\rangle $ & -- & $2806$ & $-28.3$ & $-57.6$ \\ \hline
$|2,5/2\rangle $ & $2737(12)$ & $2850$ & $15.8$ & $32.1$ \\
$|3,5/2\rangle $ & $2763.5$ & $2863$ & $28.8$ & $58.7$ \\
\hline\hline
\end{tabular}
\end{table}

The parameters for the D-wave $D$ mesons which reproduce the mass splitting
$\Delta M$(rescaled) in Table VIII are
\begin{equation}
\begin{Bmatrix}
a_{1}\text{(MeV),} & a_{2}\text{(MeV)} & b\text{(MeV)} \\
47.6, & 12.6, & 0.52, \tag{C23}
\end{Bmatrix}
\end{equation}%
where the parameter $c$ in the hyperfine term has been set to be zero. This completes the estimation of $a_{2}$ for the SH mesons in their P-wave and D-wave excitations.

\end{document}